\documentclass[aps,pra,twocolumn,superscriptaddress,nofootinbib]{revtex4-2}

\usepackage{xcolor}
\usepackage{multirow}
\usepackage{array}
\usepackage{makecell} 
\usepackage{braket}
\usepackage{graphicx}
\usepackage{amsmath}
\usepackage{amsthm}
\usepackage{amssymb}
\usepackage{hyperref}
\usepackage{enumitem}
\usepackage{float}
\usepackage[most]{tcolorbox}
\setcellgapes{3pt} 
\makegapedcells     

\usepackage{graphicx}

\newcommand{\tcell}[3]{%
  \parbox[c][#2][c]{#1}{\centering #3}%
}

\definecolor{paperblue}{RGB}{30,130,255}

\begin{document}

\title{Quantum Foundations Beyond Bell's Theorem: A Topical Review}
\author{Flavio Del Santo} 
\affiliation{University of Geneva, Group of Applied Physics, Rue de l'Ecole-de-M\'edecine 20, Geneva 1211, Switzerland}
\affiliation{Constructor University, Bremen, Germany}
\author{Nicolas Gisin} 
\affiliation{University of Geneva, Group of Applied Physics, Rue de l'Ecole-de-M\'edecine 20, Geneva 1211, Switzerland}
\affiliation{Constructor University, Bremen, Germany}

\begin{abstract}
\noindent We review a series of recent no-go theorems that extend the methodological and conceptual framework introduced by Bell's theorem to more sophisticated operational scenarios involving independent sources, quantum networks, joint measurements, and multiple observers. Many of these results were originally developed within quantum information science and are scattered across the literature, where their technical and application-driven success has partly obscured their broader foundational significance. We organize them around a common Bell-inspired methodology that extends beyond quantum nonlocality, and emphasize how increasingly rich operational structures change which physical assumptions can be tested.
\end{abstract}

\maketitle

\section{Introduction}

In 1964, John S. Bell put forward a theorem \cite{bell1964einstein, bell2004speakable} that revolutionized our understanding of quantum theory and profoundly influenced the way we think about physical reality. He showed that, assuming we are free to choose which measurements to perform, no local explanation can reproduce the predictions of quantum mechanics. Bell's theorem played a pivotal role in reopening the debate on the foundations of quantum physics and, following decades of increasingly sophisticated experimental confirmations, its impact was recognized with the 2022 Nobel Prize in Physics.

However, while Bell's theorem paved the way for modern quantum foundations, it also stimulated the systematic study of quantum entanglement and nonlocality as resources for information processing \cite{brunner2014bell}.  These lines of research ultimately gave rise to quantum information science \cite{nielsen2000quantum}. Such developments led to important technical advances like device-independent approaches, which enable the certification of quantumness solely from observed correlations, without requiring a detailed characterization of the underlying devices \cite{acin2007device,coladangelo2017all, vsupic2020self}. More broadly, quantum information has driven the development of quantum technologies \cite{acin2018quantum}, including quantum communication \cite{cleve1997substituting,gisin2007quantum}, quantum cryptography \cite{ekert1991quantum,pironio2010random,zapatero2023advances}, quantum computation \cite{proctor2025benchmarking}, and quantum metrology \cite{giovannetti2011advances}.

It seems to us that the technical and application-driven success of this research --- for example, in the thriving subfield of quantum network nonlocality \cite{cavalcanti2011quantum, branciard2012bilocal, gisin2020constraints, tavakoli2022bell} --- has, to some extent, obscured its broader conceptual significance. As research has increasingly shifted towards quantum information and its technological applications, several  fundamental advances that go beyond the lesson of Bell's theorem have, in fact, not yet been fully appreciated within quantum foundations, probably overshadowed by technical results. In particular, several recent no-go theorems extend the spirit of Bell's argument by considering scenarios involving multiple independent sources, quantum networks --- where joint measurements performed on two or more systems play an essential role --- or multiple observers.

 This paper revisits these results from a foundational perspective, emphasizing their conceptual significance rather than their technical scope. We present them as a coherent sequence of no-go theorems that naturally develops Bell’s original insight while extending beyond it.
\section{Bell scenarios}
\label{bellscenario}

We begin by recalling the Bell scenario, the basic operational framework from which the no-go theorems considered in this review naturally arise. Readers already familiar with standard Bell inequalities may therefore skip this section.

Following the standard device-independent approach (see, e.g., \cite{brunner2014bell}), the internal physical realization of the devices is left unspecified: no assumptions are made about the underlying systems, the measurement devices, or their internal functioning. The only objects of interest are the conditional probability distributions relating the observed outcomes to the chosen measurement settings --- often referred to as inputs, borrowing terminology from computer science.\footnote{Throughout this paper, unless otherwise specified, all measurements are assumed to be projective.}

\begin{figure}[]
    \centering
    \includegraphics[width=0.8\linewidth]{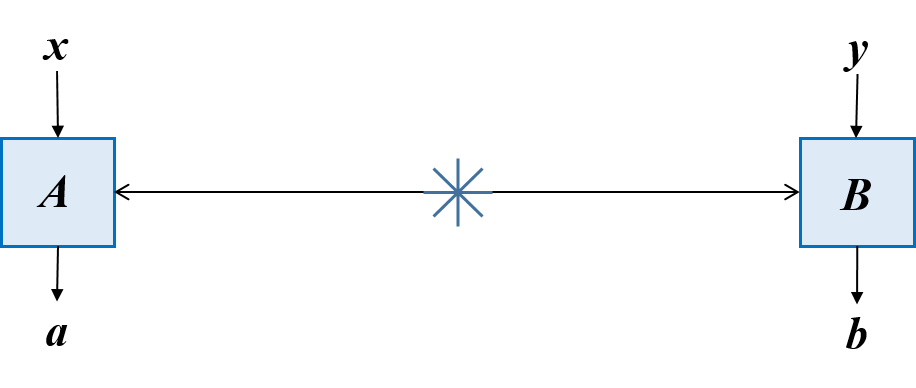}
    \caption{Bell scenario. Alice and Bob each receive a physical system from a common source. Based on their respective measurement settings $x$ and $y$, they perform local measurements on their systems and obtain outcomes $a$ and $b$, respectively. Under the assumption of locality \eqref{belllocal} , the observable correlations between the outcomes are constrained to satisfy Bell inequalities.}
    \label{bell}
\end{figure}

With reference to Fig.~\ref{bell}, consider the following Bell scenario. Two distant parties, Alice ($A$) and Bob ($B$), each receive a physical system. These systems may have previously interacted, e.g., by having been produced by a common source in their common past. In each run of the experiment, Alice chooses a measurement setting (or input) $x$ and obtains a measurement result (or outcome) $a$. Likewise, Bob chooses an input $y$ and obtains an outcome $b$.

Imagine now that the observed statistics show that
\begin{equation}
p(ab|xy)\neq p(a|x)\cdot p(b|y).
\end{equation}
This means that the outcomes obtained by Alice and Bob are correlated, and these correlations ``cry out for explanation" \cite{bell1981bertlmann}.

One possible explanation for the observed correlations is the existence of a direct causal influence between Alice and Bob.
Let us, however, suppose that the experiment is repeated under conditions that exclude any operationally detectable direct influence, i.e., under the condition of no-signaling (in both directions):
\begin{equation}
\label{ns}
\begin{aligned}
p(b| x,y) &= p(b| x',y), \\
p(a| x,y) &= p(a| x,y'),
\end{aligned}
\end{equation}
where $p(b|x,y)=\sum_a p(a,b|x,y)$, and analogously for
$p(a|x,y)$; these equalities are required to hold for all
inputs and outputs. To enforce this, for example, the two parties may be separated by a virtually impenetrable barrier (such as a sufficiently thick wall of reinforced concrete) that, to the best of our current theoretical understanding, prevents any physical medium from passing between them, thereby enforcing the no-signaling condition by excluding any physical communication. More fundamentally, one can arrange for their measurements to be spacelike separated. In that case, no signal traveling at or below the speed of light can propagate between Alice and Bob during the experiment, ruling out any direct causal explanation mediated by influences propagating within the speed of light.\footnote{One could still hypothesize direct hidden influences propagating faster than light. In Sect.~\ref{vcausalsect}, we show how models in which such influences propagate at any finite superluminal speed in a privileged reference frame can be tested against operational no-signaling.}

Within a classical causal account respecting this spacetime arrangement, the correlations must therefore be attributed to a common cause in the shared past (in accordance with Reichenbach's common cause principle \cite{sep-physics-Rpcc}).
This would mean that there are some unspecified \textit{hidden variables} $\lambda \in \Lambda$ --- hidden in the sense that they are not part of the quantum description and are not controlled in experiments, so that the observed statistics are averages over them --- which, if taken explicitly into account, would explain the observed correlations:
\begin{equation}
\label{belllocal}
    p(ab|xy)= \int_\Lambda d\lambda \ q(\lambda) p(a|x,\lambda)p(b|y,\lambda).
\end{equation}
Note that the argument does not assume determinism, since $\lambda$
may itself be a random variable distributed according to some
$q(\lambda)$: moreover the probabilities $p(a|x,\lambda)$ and $p(b|y,\lambda)$ need not be limited to the values 0 and 1. Condition \eqref{belllocal} is called \textit{locality} because the correlations are explained only in terms of common causes and local measurements.\footnote{This assumption is also referred to  as \textit{local causality} or \textit{local realism}. The latter expression is widespread but potentially misleading, since it suggests that Bell’s theorem depends on an independent and unambiguous assumption of ``realism” in addition to locality (see \cite{laudisa2023and} for a historical and philosophical discussion).} In particular, the factorization implies no-signaling for every $\lambda$, as expected for a local hidden variable.

Moreover, in the previous factorization we have implicitly assumed that the inputs $x$ and $y$ can be chosen freely, i.e., that they are statistically independent of the common cause $\lambda$:
\begin{equation}
\label{freedom}
q(\lambda|xy)=q(\lambda).
\end{equation}
This condition is called \textit{freedom of choice} (or measurement independence).\footnote{By Bayes' theorem, Eq.~\eqref{freedom}
is equivalent to $q(xy|\lambda)=q(xy)$.
Thus, knowing the hidden variable $\lambda$ does not
change the probabilities of choosing the measurement
settings $x$ and $y$, which justifies the name
\textit{freedom of choice}.} Interestingly, this can be relaxed to allow an arbitrarily small degree of measurement independence \cite{colbeck2012free,putz2014arbitrarily}, but it cannot be completely eliminated in order to derive Bell inequalities.

\begin{figure}[h!]
    \centering
    \includegraphics[width=0.6\linewidth]{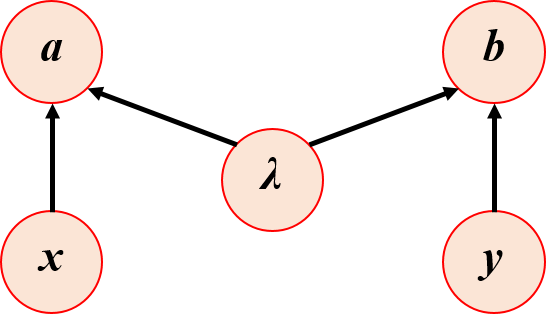}
    \caption{Causal diagram of a bipartite Bell scenario.}
    \label{bellcausal}
\end{figure}

The causal structure of a Bell scenario of the kind here described is illustrated in Fig. \ref{bellcausal}.\footnote{More generally, Bell's theorem is a special case of the problem of deciding whether given correlations are compatible with a classical causal model defined on an arbitrary directed acyclic graph. Quantum nonclassicality has been shown even in a graph with only three binary observed variables and no measurement choices~\cite{lauand2025minimal}.}

Let us consider the simplest Bell scenario, where the inputs are $x,y\in\{0,1\}$ and the outcomes $a,b\in\{-1,+1\}$. We define the correlators
\begin{equation}
\label{correlations}
\langle a_x b_y \rangle = \sum_{a,b} ab\, p(ab|xy).
\end{equation}

Assuming locality and freedom of choice, i.e., substituting Eq.~\eqref{belllocal} into Eqs.~\eqref{correlations} and combining them, one obtains
\begin{equation}
S:=\langle a_0b_0\rangle+\langle a_0b_1\rangle+\langle a_1b_0\rangle-\langle a_1b_1\rangle\le2,
\label{chsh}
\end{equation}
which is known as the Clauser--Horne--Shimony--Holt (CHSH) inequality \cite{Clauser1969}, the simplest Bell inequality. This tells us that any observed statistics from an experiment described by any local theory must satisfy this bound.
This can be generalized as follows:
\begin{tcolorbox}[
     colback=paperblue!5,
    colframe=paperblue,
    sharp corners,
    boxrule=0.5pt
]
\textbf{Bell inequalities}\\

Any theory satisfying the assumptions (i) locality (Eq.   \eqref{belllocal}) and (ii) freedom of choice (Eq. \eqref{freedom}), i.e., admitting a local hidden-variable model, must obey constraints on observable correlations, known as Bell inequalities.
\end{tcolorbox}
Note that, contrary to a widespread misconception, Bell inequalities are, strictly speaking, neither about relativity nor about quantum theory. They merely impose upper bounds on the strength of correlations classically achievable between separated parties that cannot communicate and whose measurement choices are free. 

In general, for any finite Bell scenario --- specified by fixed but arbitrary numbers of parties, measurement settings, and outcomes --- the set of correlations admitting a local hidden variable model forms a convex set with finitely many extremal points, i.e., a polytope, in the space of conditional probability distributions whose facets define tight Bell inequalities \cite{pitowsky1991correlation, werner2001all, collins2002bell, brunner2014bell}.

One may then ask whether the predictions of a given physical theory satisfy these inequalities. For quantum mechanics, the answer leads to the following \cite{bell1964einstein}:
\begin{tcolorbox}[
     colback=paperblue!5,
    colframe=paperblue,
    sharp corners,
    boxrule=0.5pt
]
\textbf{Bell's Theorem}\\

The predictions of quantum theory violate some Bell inequalities. Consequently, they cannot, in general, be reproduced by any local {(hidden-) variable} theory.
\end{tcolorbox}
\begin{proof}
 Let Alice and Bob each receive one two-level quantum system (qubit) whose joint state is maximally entangled, i.e., one of the four Bell states, e.g., $|\Psi^-\rangle
=
\frac{1}{\sqrt{2}}
\left(
|01\rangle-|10\rangle
\right)$, where $\ket{0}$ and $\ket{1}$ denote an orthonormal basis for  the qubit degree of freedom.

Let Alice and Bob each freely choose between two incompatible measurements \cite{wolf09}, with settings $x$ and $y$ corresponding to unit vectors $\vec{x}$ and $\vec{y}$ and associated with measurements of
$
\vec{x}\cdot\vec{\sigma}
$
on the first qubit and
$
\vec{y}\cdot\vec{\sigma}
$
on the second qubit, where
$
\vec{\sigma}
=
(\sigma_1,\sigma_2,\sigma_3)
$
denotes the vector of Pauli matrices.
Quantum formalism predicts  $
\langle a_xb_y\rangle
=
-\vec{x}\cdot\vec{y}.
$

If the two values of $x$ correspond to two orthogonal directions
$
\hat{ e}_1
$
and
$
\hat{ e}_2,
$
respectively, while $y\in\{0,1\}$ corresponds to measurements along the tilted directions
$
-\left(\hat{ e}_1+\hat{ e}_2\right)/\sqrt{2}
$ and $\left(-\hat{ e}_1+\hat{ e}_2\right)/\sqrt{2},
$
we obtain
$
\langle a_0b_0\rangle
=
\langle a_0b_1\rangle
=
\langle a_1b_0\rangle
=
{1}/{\sqrt{2}}$ 
and $\langle a_1b_1\rangle
=
-{1}/{\sqrt{2}}.
$

Therefore,
\begin{equation}
S=2\sqrt{2},
\end{equation}
which violates the CHSH inequality given in Eq. \eqref{chsh}. This shows that the predictions of quantum mechanics cannot, in general, be accounted for by local hidden-variable models. 
\end{proof}
Different interpretations of quantum theory provide different accounts of the consequences of Bell's theorem (see, e.g., \cite{cabello2017interpretations}), but no single explanation of quantum nonlocality has been universally accepted to date.

Violations of Bell inequalities have been experimentally
observed over several decades in increasingly ambitious
experiments \cite{aspect1981experimental,tittel1998violation,
tittel1999long}, culminating in loophole-free Bell tests
that closed all major experimental loopholes
\cite{hensen2015loophole,shalm2015strong,
giustina2015significant, rosenfeld2017event}. Hence, beyond any reasonable doubt, nature cannot be described by any local hidden-variable model, and any future physical theory, even one extending beyond quantum mechanics, will have to account for this fact.

\section{No-go theorems beyond Bell}

A no-go theorem is a formal argument showing that a set of physically motivated assumptions is jointly incompatible with certain predictions of a theory --- here, typically, observable correlations predicted by quantum theory --- so that at least one of the assumptions must be abandoned. As already recalled, Bell's (no-go) theorem showed that operational constraints can rule out entire classes of ontological explanations, thereby giving rise to what Shimony referred to as \textit{experimental metaphysics} \cite{shimony1989search}. Over the past two decades, this style of reasoning has expanded far beyond the original Bell scenario, yielding no-go theorems that constrain the reality of the quantum state, admissible causal structures, observer-independent facts, and even the mathematical field over which quantum theory is formulated.\footnote{Another major line of research concerns contextuality, inaugurated by the Bell--Kochen--Specker no-go theorem \cite{bell1966problem,kochen2011problem}. We  omit this vast subject here, as our focus is instead on Bell-inspired no-go theorems arising from extended operational scenarios involving multiple physical systems, distributed among several parties.} Without any pretense of completeness, we review here a selection of these developments, some of which have received comparatively little attention in the quantum foundations literature, and, where possible, provide intuitive, non-technical sketches of the underlying proofs. Our aim is to show that Bell's theorem did more than rule out local hidden-variable theories: it laid the foundations for a new way of investigating the limits of physical theories through operational impossibility results.
\subsection{Bell nonlocality without measurement choices}
\label{bellnosettings}

Bell's theorem relies on the possibility to choose freely among different measurement settings. However, an important conceptual advancement from quantum networks is the realization that this assumption can be replaced by the independence of several physical sources, making it possible to demonstrate nonlocality even when every observer performs a single fixed measurement \cite{branciard2012bilocal, fritz2012beyond,gisin2020constraints}.

\begin{figure}[h!]
    \centering
    \includegraphics[width=0.8\linewidth]{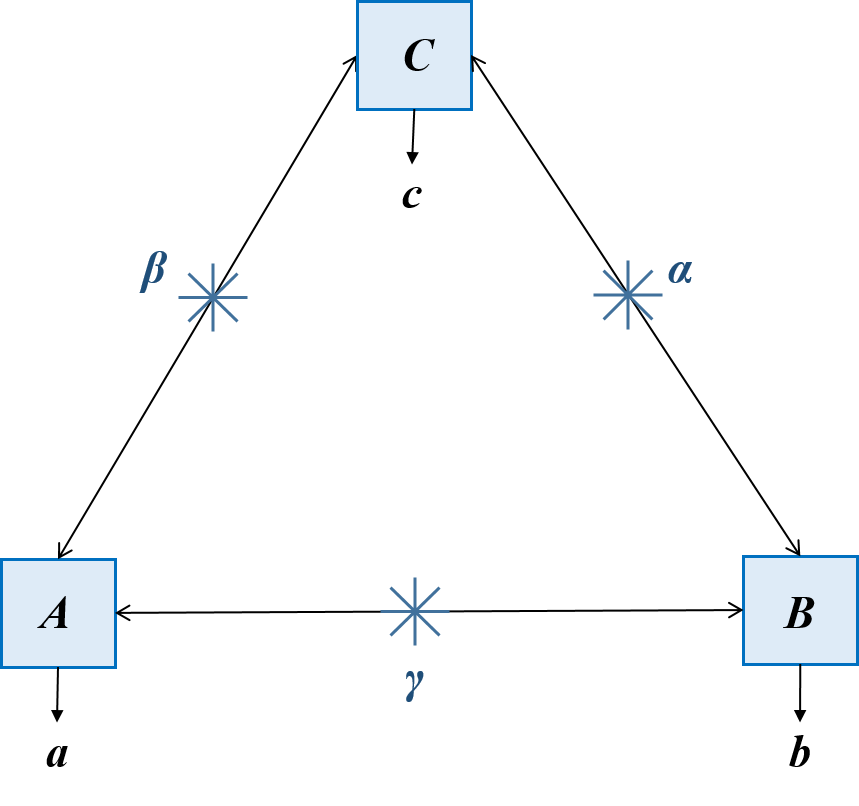}
    \caption{Triangle network. Three independent sources distribute entangled pairs to Alice, Bob, and Charlie, who obtain outputs $a$, $b$, and $c$ from fixed joint measurements on the systems they receive.}
    \label{triangle}
\end{figure}

Consider now the simplest example in terms of the number of parties,\footnote{Note that a linear network without inputs requires at least four parties \cite{fritz2012beyond, don2026minimal}. The first example of nonlocality without inputs was indeed given in such a four-party linear network~\cite{branciard2012bilocal}. The network that demonstrates nonlocality without inputs with the smallest number of parties is, however, the triangle.} i.e., the triangle network depicted in Fig.~\ref{triangle}. Three independent sources distribute physical systems to the pairs of observers $AB$, $BC$, and $CA$. Alice, Bob, and Charlie perform fixed joint measurements on the systems they receive and obtain outcomes $a$, $b$, and $c$, respectively.

A classical model envisions three independent local hidden variables $\alpha$, $\beta$, and $\gamma$, each associated with one of the independent sources, that account for the causal structure. The outcome distribution then has the form: 
\begin{equation}
\begin{aligned}
p(a,b,c)
={}&
\int d\alpha\,d\beta\,d\gamma\,
q_\alpha(\alpha)q_\beta(\beta)q_\gamma(\gamma)
\\
& p(a|\beta,\gamma)\,
p(b|\gamma,\alpha)\,
p(c|\alpha,\beta).
\end{aligned}
\end{equation}
The factorization of the hidden-variable distributions expresses the assumption that the three sources are statistically independent. Correlations admitting such a decomposition are called \emph{triangle-local}.

In Refs.~\cite{fritz2012beyond,renou2019genuine}, the following result was established:

\begin{tcolorbox}[
    colback=paperblue!5,
    colframe=paperblue,
    sharp corners,
    boxrule=0.5pt
]
\textbf{Nonlocality without inputs theorem}\\

Any hidden-variable model satisfying locality and statistical independence of the sources fails to reproduce certain correlations predicted by quantum theory in the triangle network. Consequently, freely chosen measurement settings are not necessary to demonstrate quantum nonlocality.
\end{tcolorbox}
\noindent Proofs of network nonlocality are generally difficult because the set of triangle-local correlations is non-convex (see, e.g., \cite{gisin2019entanglement, baumer2025exploring}). A powerful approach is provided by the inflation technique, which constructs enlarged causal structures containing several copies of the original variables and derives consistency conditions that must be satisfied by every distribution
compatible with the original network \cite{wolfe2019inflation}. It would nevertheless be desirable to develop further techniques for proving network nonlocality (see e.g., \cite{girardin2026versatile, don2026minimal, dasilvalocal}), which remains a difficult task even for a single fixed choice of parameters, let alone for characterizing the full nonlocal set.

As an illustrative example, let each source independently distribute a
singlet state $\ket{\Psi^-}$, and let each party perform the
\textit{Elegant Joint Measurement} introduced in
Ref.~\cite{gisin2019entanglement}.  This is a projective measurement onto
four partially entangled two-qubit states whose local reduced states,
represented by their Bloch vectors, form regular tetrahedra on the Bloch
sphere. The resulting highly symmetric distribution was conjectured for
several years to be triangle-nonlocal. Gitton and Renner recently proved
this conjecture by combining inflation techniques with symmetry reductions
and computer-assisted optimization carried out in exact arithmetic
\cite{gitton2025elegant}.

Interestingly, if the three independent sources distribute Bell states
and each observer performs a Bell-state measurement, the resulting
correlations admit a triangle-local model \cite{gisin2019entanglement}. Thus, quite counterintuitively, maximal entanglement of both the source states and the measurement eigenstates --- i.e., the seemingly most non-classical scenario --- provably does not lead to triangle nonlocality. More generally, it was recently shown  that independently of the dimension, if all three sources in the triangle network distribute maximally entangled states and all parties perform projective maximally entangled measurements, then the resulting distribution admits a triangle-local model (see Appendix G of \cite{svegborn2026uncountably}).  This shows that the relation between entanglement and nonlocality in networks depends on a subtler interplay between the distributed states and the joint measurements.

More generally, no-signaling and source independence alone impose nontrivial constraints on the correlations achievable in the triangle network, despite the absence of measurement inputs \cite{gisin2020constraints}. Remarkably, the demonstration of network nonlocality does not require perfect source independence: it can still be obtained assuming only an arbitrarily small degree of independence among the sources \cite{supic2020quantum}. However, if experimental rounds are not assumed to be independent and identically distributed, the nonclassicality of networks without inputs cannot be certified: parties exploiting memory across rounds and minimal prior coordination can mimic a common source of shared randomness \cite{weilenmann2025memory}. This unclosable loophole can be regarded as a network analogue of \textit{superdeterminism}, since it undermines source independence, which here plays the role of freedom of choice.

Interestingly, in standard Bell scenarios, demonstrating nonlocality requires the parties to choose between incompatible measurements, making the role of measurement incompatibility --- and hence of Planck's reduced constant $\hbar$ --- explicit in the emergence of non-classical correlations. In network scenarios without inputs, by contrast, this connection is more subtle, since the relevant non-classical structure must arise from the interplay between the fixed measurements and the independent states distributed by the sources. 

In this context, it has been shown that coherence between different modes of a single particle can already give rise to nonlocal correlations in triangle networks with fixed measurements, and that such coherence can itself be certified without requiring measurement inputs \cite{abiuso2022single,bibak2024quantum}.

Network nonlocality is not restricted to the triangle scenario.
Bilocal correlations and their quantum violations were first characterized in entanglement-swapping networks \cite{branciard2012bilocal}, and related
constraints were later derived for star-shaped networks involving several independent sources \cite{tavakoli2014correlations}. Moreover,
connected networks of bipartite pure entangled states were shown to generate genuinely multipartite nonlocality
\cite{contrerastejada2021genuine}, while the stronger notions of full network nonlocality and genuine network nonlocality were motivated by the search for correlations that cannot be reproduced by embedding pairwise Bell scenarios within a network \cite{pozaskerstjens2022full, supic2020quantum}; a criterion specifically designed to certify network nonclassicality not traceable to embedded Bell scenarios was later proposed in Ref.~\cite{ciudad2026escaping}.

In general, at the conceptual level, quantum networks shift the focus beyond entangled states to joint entangled measurements, which are still in the process of being studied systematically \cite{chitambar2014everything, del2024iso, pauwels2025classification}. Understanding joint measurements at a fundamental level is expected to provide new insights into the foundations of quantum nonlocality \cite{cavalcanti2023fresh}.


\subsection{No finite-speed superluminal explanation of Bell nonlocality}
\label{vcausalsect}

In Sect. \ref{bellscenario}, we recalled that Bell tests are typically performed with spacelike-separated measurements, so that relativity excludes ordinary subluminal communication between the parties. It is therefore the validity of special relativity that allows one to exclude direct causal influences between Alice and Bob.

\begin{figure}[h]
    \centering
    \includegraphics[width=\linewidth]{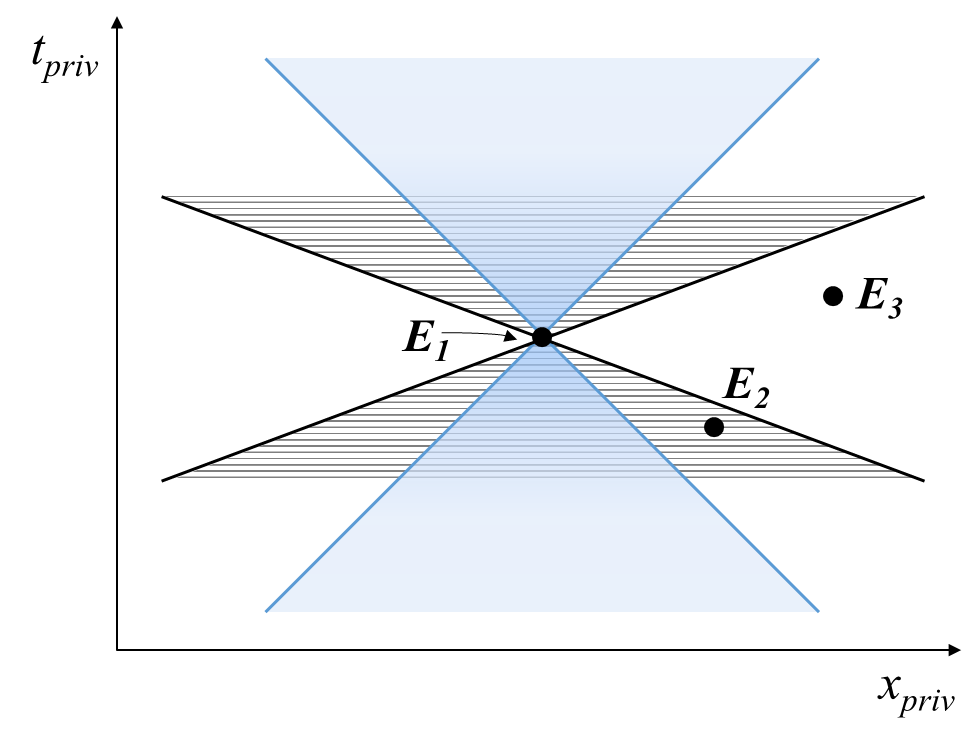}
    \caption{Space-time diagram comparing the light-cone (in light blue) and the $v$-cone (hatched) in the privileged reference frame (with coordinates indicated by subscript \textit{priv}). Both $E_2$ and $E_3$ are spacelike separated from $E_1$. However, $E_1$ and $E_2$ are $v$-causally connected, meaning that hidden influences at speed $v$ can transfer from $E_2$ to $E_1$.}
    \label{vcone}
\end{figure}

In principle, however, special relativity would be empirically compatible with the existence of a hypothetical privileged reference frame in which influences can travel at a finite but arbitrarily large speed $v > c$, as long as these are not detectable, otherwise, this would falsify relativity. This can be represented by associating with an event $E_1$ not only the usual light cone, within which detectable causal influences can propagate, but also a $v$-causal cone, within which hidden superluminal influences can propagate. This is illustrated in Fig. \ref{vcone}, where $E_2$ is spacelike separated from $E_1$, but it lies in its $v$-causal past. By contrast, $E_3$ lies outside both the light-cone and the $v$-cone of $E_1$ and therefore cannot exchange any signal
with it, even if hidden.

These ideas were put forward in a series of works \cite{eberhard1989realistic, scarani2000speed, scarani2002superluminal, scarani2005superluminal,  bancal2013quantum, barnea2013tripartite,  gisin2014quantum}, where the authors defined \emph{operationally no-signaling $v$-causal models}, namely, models in which correlations are explained by a combination of hidden common causes and direct causal influences propagating at a finite but superluminal speed $v>c$ in a hypothetical privileged reference frame.

\begin{figure}[h!]
    \centering
    \includegraphics[width=0.6\linewidth]{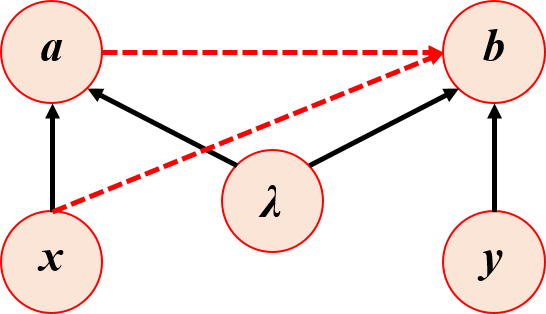}
    \caption{Causal diagram of a bipartite Bell scenario with possible
(hidden) superluminal influences (dashed red arrows). These are assumed to propagate at a finite but arbitrarily large speed $v>c$  with respect to a hypothetical preferred foliation of spacetime. The diagram corresponds to the case $A\prec B$, i.e., where Alice's measurement precedes Bob's in the privileged reference frame.}
    \label{vcausalcausal}
\end{figure}

Consider now a bipartite Bell scenario, as described in Sect. \ref{bellscenario}. Assume that Alice lies in the $v$-causal past of Bob, denoted by $A\prec B$. Then hidden influences emitted during Alice's measurement may reach Bob before he performs his own measurement. Consequently, Bob's outcome may depend on Alice's local measurement choice and outcome. Hence, the joint probability distribution of the outcomes $a, b$ given the inputs $x,y$ can be decomposed as:
\begin{equation}
    p^{A \prec B}(a,b| x,y)=\int_{\Lambda} d\lambda\,q(\lambda)\,p(a| x,\lambda)\,p(b| a,x,y,\lambda).
\end{equation}
    This corresponds to the causal diagram depicted in Fig. \ref{vcausalcausal}.

It is possible to show that, in the bipartite case, any quantum correlation can be reproduced by a suitable $v$-causal model, while the underlying superluminal influences remain operationally undetectable \cite{bancal2013quantum}. One might therefore conclude that no empirical test can rule out finite-speed hidden influences as a viable explanation of Bell nonlocality. Surprisingly, this turns out to be true only in the bipartite case.

By contrast, it is possible to prove that there exist three-partite \cite{barnea2013tripartite} and four-partite \cite{bancal2013quantum} scenarios which allow $v$-causal explanations to be tested against operational no-signaling.\footnote{Tripartite tests can also exploit the fact that nonlocality can be transitive: nonlocal correlations between $A$ and $B$ and between $B$ and $C$ can force those between $A$ and $C$ to be nonlocal too \cite{coretti2011nonlocality}. This was recently  shown also for quantum states  \cite{chen2026nonlocality}.} This leads to the following general theorem:


\begin{tcolorbox}[
     colback=paperblue!5,
    colframe=paperblue,
    sharp corners,
    boxrule=0.5pt
]
\textbf{Finite-speed superluminal-influence no-go theorem}\\

For every finite speed $v$, no $v$-causal model can reproduce all quantum
correlations while remaining operationally no-signaling. Hence, any
explanation of quantum correlations in terms of causal influences
propagating at a finite (superluminal) speed in a privileged reference frame
necessarily predicts detectable signaling traveling at $v$.
\end{tcolorbox}
\begin{proof}
We now sketch the original four-party argument of Ref.~\cite{bancal2013quantum}, which is more intuitive than the three-party construction in Ref.~\cite{barnea2013tripartite}. The argument aims at ruling out a $v$-causal explanation of Bell nonlocality. It retains the freedom-of-choice assumption and requires operational no-signaling.

With reference to Fig.~\ref{vcausalproof}, let Alice, Bob, Charlie, and Debbie be arranged, in this hypothetical privileged reference frame, according to the causal ordering
\begin{equation}
A\prec D\prec(B\sim C),
\end{equation}
where $B\sim C$ means that Bob's and Charlie's measurements are quasi-simultaneous so that neither event lies in the other's $v$-cone. Alice, Bob, Charlie, and Debbie freely choose inputs $x,y,z,w\in\{0,1\}$, respectively, and obtain outcomes $a,b,c,d\in\{-1,+1\}$.

\begin{figure}[h!]
    \centering
    \includegraphics[width=0.9\linewidth]{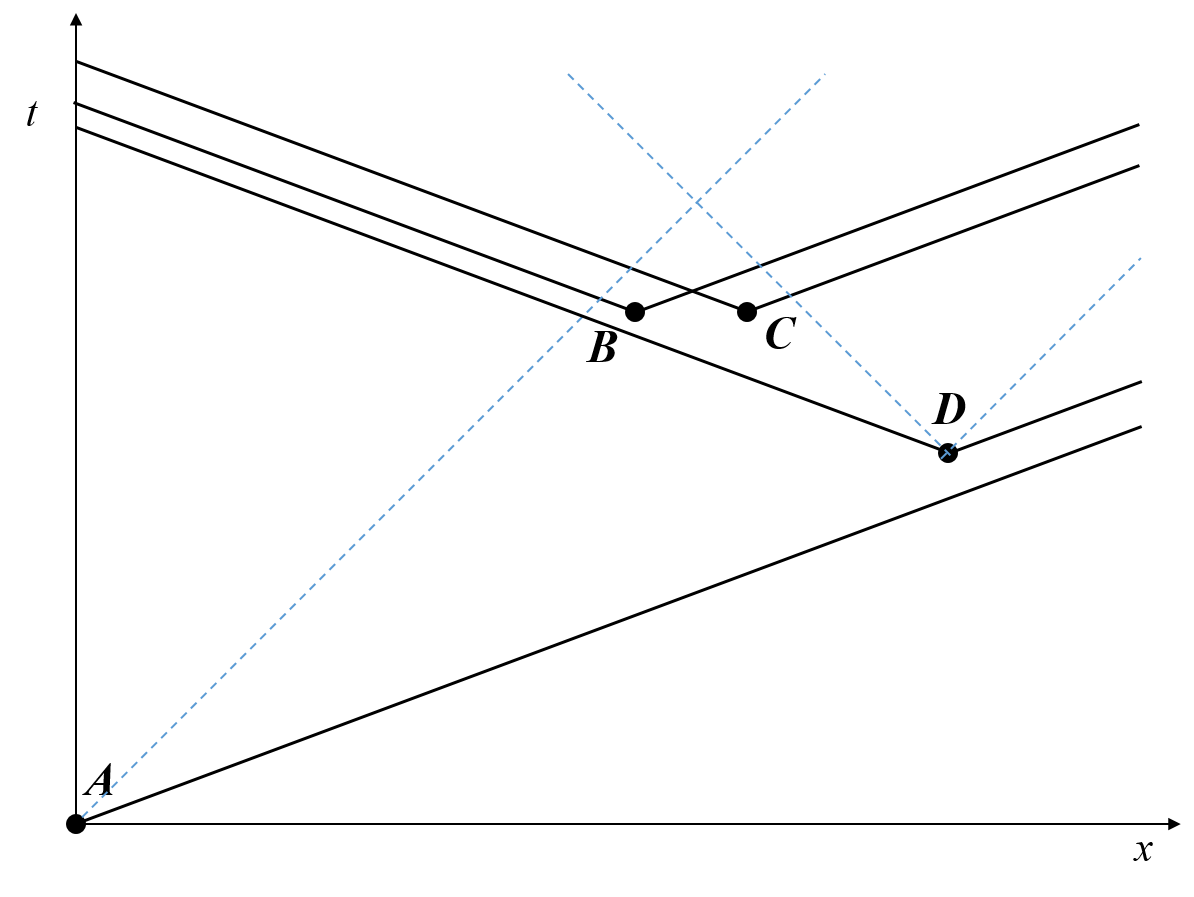}
    \caption{Space-time diagram ($v$-cones represented in black, light-cones dashed in blue) in the hypothetical privileged reference frame of the four-partite scenario that shows that $v$-causal models lead to detectable signaling.}
    \label{vcausalproof}
\end{figure}

Since no $v$-causal direct influence can travel between Bob and Charlie, their correlations may only arise from  common causes --- denoted by the hidden variable $\mu$ --- and from information previously propagated from Alice and Debbie who lie in their past $v$-cones. Thus, conditioned on Alice's and Debbie's inputs and outputs, the correlations between Bob and Charlie are local (cf. Eq. \eqref{belllocal}):

\begin{equation}
\begin{aligned}
p(b,c|y,z,a,x,d,w)
&=
\int d\mu\,
q(\mu|a,x,d,w)\cdot\\
&\qquad
p(b|y,\mu)\cdot
p(c|z,\mu).
\end{aligned}
\end{equation}

Moreover, despite the hypothesized superluminal hidden influences, we require that they cannot give rise to detectable superluminal signaling. This is expressed by the \textit{operational no-signaling} condition
\begin{equation}
\sum_d p(a,b,c,d|x,y,z,w)
=
p(a,b,c|x,y,z),
\end{equation}
together with the three analogous conditions for the $ABD$, $ACD$, and $BCD$ marginals, obtained by marginalizing over the outcome of the remaining party.

Bancal et al.~\cite{bancal2013quantum} showed that the conjunction of the two previous assumptions --- i.e., (i) conditional Bell locality of Bob and Charlie and (ii) operational no-signaling --- implies the following Bell-like inequality:
\begin{equation}
    \label{vcausalineq}
\begin{aligned}
S := {}&
-3\langle a_0\rangle
-\langle b_0\rangle
-\langle b_1\rangle
-\langle c_0\rangle
-3\langle d_0\rangle\\
&-\langle a_1b_0\rangle
-\langle a_1b_1\rangle
+\langle a_0c_0\rangle
+2\langle a_1c_0\rangle\\
&+\langle a_0d_0\rangle
+\langle b_0d_1\rangle
-\langle b_1d_1\rangle
-\langle c_0d_0\rangle\\
&-2\langle c_1d_1\rangle
+\langle a_0b_0d_0\rangle
+\langle a_0b_0d_1\rangle
+\langle a_0b_1d_0\rangle\\
&-\langle a_0b_1d_1\rangle
-\langle a_1b_0d_0\rangle
-\langle a_1b_1d_0\rangle
+\langle a_0c_0d_0\rangle\\
&+2\langle a_1c_0d_0\rangle
-2\langle a_0c_1d_1\rangle
\leq 7,
\end{aligned}
\end{equation}
where, for example, $\langle a_x\rangle$ denotes the expectation value of Alice's outcome for input $x$, while $\langle a_xb_y\rangle$ and $\langle a_xb_yd_w\rangle$ denote the corresponding two- and three-party correlators, respectively, and analogously for all the other terms.\footnote{A considerably simpler inequality, involving only six correlators and with a larger relative quantum violation, has recently been derived in Ref.~\cite{li2026clusterstatewitnessesfinitespeedhidden}.}

However, for a suitable four-qubit entangled state and local binary measurements, quantum theory predicts a violation of this inequality  \cite{bancal2013quantum}:
\begin{equation}
\label{sviol}
S\simeq 7.2>7.
\end{equation}

To test a possible violation of inequality \eqref{vcausalineq}, a challenge remains. Since the speed $v$ and the hypothetical privileged reference frame are unknown, it is not obvious how to guarantee the configuration of Fig. \ref{vcausalproof}. In fact, for $A\prec D\prec B$ and $A \prec D \prec C$ one may merely use the light cone; since $c<v$, this guarantees the required orderings. Yet nothing of the sort can be done to ensure $B\sim C$. However, the inequality in Eq. \eqref{vcausalineq} has the crucial property that none of its terms contains both Bob's and Charlie's outcomes: it depends only on the $ABD$ and $ACD$ marginals. This special structure is what allows one to prove the theorem.

To see this, suppose that Charlie delays his measurement so that it lies in the future $v$-cone of Bob's measurement. In this ordering, a $v$-causal model reproducing the quantum correlations must reproduce, in particular, the quantum $ABD$ marginal. But since delaying Charlie cannot affect the statistics of Alice, Bob, and Debbie, the same $ABD$ marginal must also be obtained in the original configuration $B\sim C$.

By the same reasoning, delaying Bob's measurement into the future $v$-cone of Charlie's measurement event shows that the $ACD$ marginal in the original configuration must likewise coincide with its quantum prediction.

Thus, both marginals entering $S$ retain their quantum values, and the $v$-causal model must reproduce the quantum value of $S$ also in the original configuration $B\sim C$. Yet, in this configuration, the correlations between Bob and Charlie, conditioned on Alice's and Debbie's settings and outcomes, are local by construction. Together with operational no-signaling, this is precisely the condition under which Bancal \textit{et al.}'s inequality requires $S\leq 7$. Therefore, the $v$-causal model must necessarily violate the operational no-signaling condition, thereby leading to detectable superluminal signaling, and hence to the exclusion of hidden $v$-causal models or of relativity. 
\end{proof}

This theorem has not yet been experimentally tested, either in the tripartite scenario of Ref. \cite{barnea2013tripartite} or in the four-partite scenario of Ref. \cite{bancal2013quantum}. An experimental test would involve observing a violation of an inequality such as \eqref{vcausalineq}, which appears to be within experimental reach. Such a result would have significant foundational implications: an observed violation, together with operational no-signaling, would rule out any explanation with $v$-causal models. Conversely, maintaining a \(v\)-causal explanation would require detectable faster-than-light signaling, and hence a conflict with relativity.

However, several bipartite experiments have been already conducted, which placed lower bounds on the speed of hypothetical superluminal influences underlying quantum correlations. Salart \textit{et al.}~\cite{salart2008testing}, using entangled photons distributed over $18\,\mathrm{km}$ and exploiting the Earth's rotation to probe different possible privileged frames, found that such an influence would have to propagate at least about $10^4 \text{–} 10^5$ times faster than the speed of light, depending on the orientation of the hypothetical privileged frame and the Earth's velocity relative to it. Cocciaro \textit{et al.}~\cite{cocciaro2011lower} obtained a comparable bound, $v\gtrsim 0.6\times10^4c$, for privileged frames moving at less than $0.1c$ with respect to the Earth.

Subsequent experiments further strengthened these tests. Yin \textit{et al.}~\cite{yin2013lower} addressed the locality and freedom-of-choice loopholes while retaining a lower bound of order $10^4c$, assuming the Earth's speed in any inertial reference frame to be smaller than $10^{-3}c$. Cocciaro \textit{et al.}~\cite{cocciaro2018improved} instead optimized the sensitivity to the propagation speed, obtaining bounds as high as $5\times10^6c$ for preferred-frame velocities comparable to that of the cosmic-microwave-background frame. Moreover, meter-scale tabletop implementations have shown that bounds of order $10^4c$ can be obtained with much shorter separation distances and acquisition times \cite{santamaria2023testing}. Very recently, more experimentally accessible multipartite tests based on linear cluster states have been proposed \cite{li2026clusterstatewitnessesfinitespeedhidden}.

 Finally, the argument of Ref.~\cite{bancal2013quantum} applies more generally to any hidden-influence model that reproduces the quantum correlations in some configurations while yielding only local correlations in others, such as models in which entanglement disappears beyond a certain distance, or the ``before-before'' models \cite{scarani2014strong}.

\subsection{The reality of the quantum state}

We have seen the limitations of local (even hypothetically
superluminal) hidden-variable explanations of quantum
predictions. A distinct but related question is whether the
quantum state has an objective physical status, or merely
represents incomplete knowledge about a hidden ontic state.

The crucial question is whether the underlying ontic state
$\lambda$ uniquely determines the quantum state.
Consider two distinct pure states $\ket{\psi_1}$ and
$\ket{\psi_2}$, associated with probability distributions
$\mu_{\psi_1}(\lambda)$ and $\mu_{\psi_2}(\lambda)$ over
the hypothetical ontic state space.

A model is called \emph{$\psi$-ontic} if, for every pair of
distinct pure states, the corresponding distributions have
disjoint supports ~\cite{harrigan2010einstein}. In other words, there is no ontic state
$\lambda$ that can occur with nonzero probability under
both preparations. Hence, knowing $\lambda$ uniquely
determines which quantum state was prepared.

Conversely, a model is called \emph{$\psi$-epistemic} if there exist two distinct pure states for which the corresponding ontic-state distributions overlap. More precisely, there exists a region $\Delta$ of the ontic state space that lies
in the support of both $\mu_{\psi_1}$ and $\mu_{\psi_2}$. Hence, an ontic state $\lambda \in \Delta$ is compatible with
either preparation.

Therefore, if $\lambda\in\Delta$, the same underlying
physical state is compatible with either preparation,
$\ket{\psi_1}$ or $\ket{\psi_2}$. The quantum state is
then not uniquely determined by the underlying physical
state and can be interpreted as encoding incomplete
information about it.

\begin{figure}[h!]
    \centering
    \includegraphics[width=0.8\linewidth]{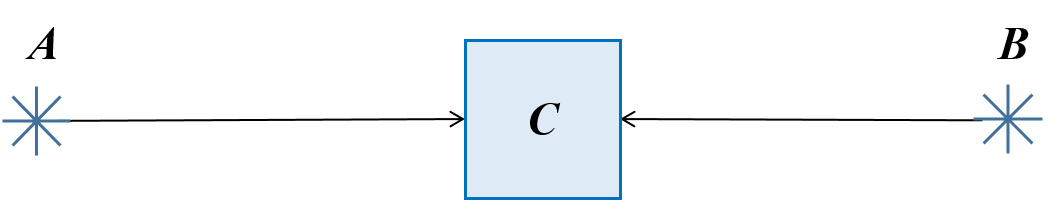}
    \caption{PBR scenario. Alice and Bob each independently
    prepare one of two nonorthogonal states, represented
    in their respective local Hilbert spaces by $\ket{\psi_1}$ or
    $\ket{\psi_2}$ for Alice and $\ket{\phi_1}$ or
    $\ket{\phi_2}$ for Bob. The two systems are sent
    towards Charlie, who performs a standard Bell-state
    measurement (see main text).}
    \label{pbr}
\end{figure}

Consider now two independently prepared systems, the first
of which may be prepared in either $\ket{\psi_1}$ or
$\ket{\psi_2}$, and the second in either $\ket{\phi_1}$
or $\ket{\phi_2}$.

The key assumption here is \emph{preparation independence},
namely that independently prepared systems correspond to
statistically independent ontic states,
\begin{equation}
\mu_{\psi_i\otimes\phi_j}(\lambda_1,\lambda_2)
=
\mu_{\psi_i}(\lambda_1)\,
\mu_{\phi_j}(\lambda_2).
\end{equation}

Pusey, Barrett and Rudolph then proved the following result \cite{pusey2012reality}:

\begin{tcolorbox}[
    colback=paperblue!5,
    colframe=paperblue,
    sharp corners,
    boxrule=0.5pt
]
\textbf{PBR theorem}\\

The predictions of quantum theory are incompatible with
$\psi$-epistemic models, assuming preparation independence.
Consequently, within the ontological-models framework and
under preparation independence, the quantum state cannot
be interpreted as merely epistemic information about an
underlying ontic state.
\end{tcolorbox}

\begin{proof}
With reference to Fig.~\ref{pbr}, we give a sketch of
the original two-qubit construction of
Ref.~\cite{pusey2012reality}. Two independent sources
each prepare one of the same two nonorthogonal states,
$\ket{0}$ or $\ket{+}:=(\ket{0}+\ket{1})/\sqrt{2}$,
which are subsequently subjected to a joint measurement
in a maximally entangled basis. Since this basis is
locally unitarily equivalent to the standard Bell basis,
we present an equivalent version of the proof using
the familiar Bell-state measurement, with the preparation
states expressed in the corresponding local bases.

In these local bases, the first source prepares either
\begin{equation}
\begin{aligned}
\ket{\psi_1} &= \cos\theta\ket{0}-\sin\theta\ket{1},\\
\ket{\psi_2} &= \cos\theta\ket{0}+\sin\theta\ket{1},
\end{aligned}
\end{equation}
while the second source prepares either
\begin{equation}
\begin{aligned}
\ket{\phi_1} &= \cos\theta\ket{0}+\sin\theta\ket{1},\\
\ket{\phi_2} &= \sin\theta\ket{0}+\cos\theta\ket{1}.
\end{aligned}
\end{equation}
To convey the structure of the formal proof, we present here only the special case $\theta=\pi/8$, for which the two pairs of states have the same overlap, $\langle\psi_1|\psi_2\rangle=\langle\phi_1|\phi_2\rangle=1/\sqrt{2}$.
However, Ref.~\cite{pusey2012reality} proves the result in full generality by considering a sufficiently large number of independent preparations: for any pair of distinct pure states, any nonzero overlap of the corresponding ontic distributions leads to a contradiction with quantum theory.

Assume now, by contradiction, that the ontological model
is $\psi$-epistemic for these two preparation procedures, so that
their ontic distributions overlap: there is a set of ontic states
to which both procedures assign nonzero probability.
Since the two independent sources implement the same two
preparation procedures (in both cases the states are $\ket{0}$ and
$\ket{+}$, written in two different local bases), each source has
such an overlap region in its own ontic state space, $\Delta_\psi$
for the first and $\Delta_\phi$ for the second, to which both of
its preparations assign nonzero probability.
Accordingly, one can choose $q>0$ such that, regardless
of which state is prepared by either source, the
probability of obtaining an ontic state in the respective
overlap region is at least $q$.
For two independently prepared systems, preparation
independence implies that the probability that
$\lambda_1\in\Delta_\psi$ and $\lambda_2\in\Delta_\phi$
is at least $q^2>0$.

Since any ontic state in the respective overlap region
is compatible with either quantum state prepared by
that source, the same pair of ontic states
$(\lambda_1,\lambda_2)\in\Delta_\psi\times\Delta_\phi$
is compatible with all four possible preparations:
\begin{equation}
\begin{aligned}
\ket{\chi_1} &= \ket{\psi_1}\ket{\phi_1},\\
\ket{\chi_2} &= \ket{\psi_1}\ket{\phi_2},\\
\ket{\chi_3} &= \ket{\psi_2}\ket{\phi_1},\\
\ket{\chi_4} &= \ket{\psi_2}\ket{\phi_2}.
\end{aligned}
\label{prepar}
\end{equation}

The two independently prepared states are brought
together at $C$, where they are jointly measured
in the standard Bell basis:
\begin{equation}
\begin{aligned}
\ket{\xi_1} &= \ket{\Psi^+}
=\frac{1}{\sqrt{2}}
\left(\ket{0}\ket{1}
+\ket{1}\ket{0}\right),\\
\ket{\xi_2} &= \ket{\Phi^+}
=\frac{1}{\sqrt{2}}
\left(\ket{0}\ket{0}
+\ket{1}\ket{1}\right),\\
\ket{\xi_3} &= \ket{\Psi^-}
=\frac{1}{\sqrt{2}}
\left(\ket{0}\ket{1}
-\ket{1}\ket{0}\right),\\
\ket{\xi_4} &= \ket{\Phi^-}
=\frac{1}{\sqrt{2}}
\left(\ket{0}\ket{0}
-\ket{1}\ket{1}\right).
\end{aligned}
\end{equation}

The measurement yields an outcome
$c\in\{1,2,3,4\}$, each value corresponding to one
of the four basis states above.

It is straightforward to verify that, for each prepared
state in Eq.~\eqref{prepar}, there exists a measurement
basis state orthogonal to it, namely
$\braket{\xi_i|\chi_i}=0$,
$\forall i\in\{1,2,3,4\}$.
Hence, quantum mechanics predicts that outcome $i$
occurs with zero probability when the state
$\ket{\chi_i}$ is prepared.
However, with probability at least $q^2>0$, the two ontic
states lie in their respective overlap regions
$\Delta_\psi$ and $\Delta_\phi$, so that the same ontic
state pair $(\lambda_1,\lambda_2)$ is compatible with
all four preparations
$\ket{\chi_1},\ldots,\ket{\chi_4}$.

For ontic states in this overlap region, the measurement must yield one of the four possible outcomes. Since the region has nonzero probability, at least one outcome, say $j$, must occur with nonzero probability for a set of ontic states compatible with all four preparations. Since the outcome probabilities depend
only on the ontic state and the measurement performed,
the same ontic state pair is also compatible with the
preparation $\ket{\chi_j}$, for which the ontological
model would therefore assign a nonzero probability
to outcome $j$. 
Quantum mechanics, however, assigns zero probability
to precisely that outcome.

This contradiction shows that the ontic distributions
associated with the two alternative preparation procedures
cannot overlap. Hence, no $\psi$-epistemic model
satisfying preparation independence can reproduce the
quantum predictions for these states.
\end{proof}

The authors of the PBR theorem acknowledge that their
result is in the same spirit as Bell's theorem, as both assume the existence of an underlying physical variable $\lambda$ that determines the probabilities of measurement outcomes \cite{pusey2012reality}.
 However, they emphasize a fundamental difference between the assumptions underlying the two results: while Bell's theorem relies on freedom of measurement choice, the PBR
theorem relies instead on preparation independence.

Preparation independence is, however, a stronger requirement than it may first appear. It demands not only that the two systems be prepared independently, but also that their ontic states carry no ``instructions'' on how to behave should they later meet another system in a joint measurement. Indeed, once preparation independence is dropped, explicit $\psi$-epistemic models reproducing the quantum predictions can be constructed \cite{lewis2012distinct}. Nevertheless, it has been shown that $\psi$-epistemic models are arbitrarily bad at explaining the indistinguishability of nonorthogonal quantum states in Hilbert spaces of dimension $d\geq4$ \cite{branciard2014psi}. More recently, the impossibility of simulating a qubit channel with finite classical communication, when the transmitted system can be jointly measured with other quantum systems held by the receiver, has been argued to support the ontic status of the qubit state without invoking preparation independence, relying instead on locality constraints on the underlying ontic state space \cite{gopalkrishna2026no}.

Interestingly, this distinction becomes less sharp when one considers multipartite networks involving several sources. In Sect.~\ref{bellnosettings}, we have indeed reviewed recent results showing that Bell nonlocality can likewise be derived under the assumption of preparation independence,
without requiring any choice of measurement setting. In fact, measurement settings can be explicitly replaced by additional independent sources and parties, thereby turning freedom of choice into an assumption of source independence \cite{fritz2012beyond}.

Colbeck and Renner showed that under their freedom-of-choice assumption, and without assuming preparation independence, the underlying ontic state uniquely determines the quantum state \cite{PhysRevLett.108.150402,Colbeck_2017}. Hence, $\psi$-ontology follows from free choice without preparation independence, in a sense dual to the derivation of Bell nonlocality from source independence without freely chosen measurement settings discussed above.


\subsection{No better predictions beyond quantum theory}
\label{CR}

The PBR theorem constrains epistemic interpretations of the
quantum state, but does not exclude the possibility that the
quantum state might be supplemented by additional information.
Could such information allow us to predict measurement outcomes
more accurately than quantum theory itself? This question was
addressed by Colbeck and Renner, who derived a no-go theorem
constraining the predictive power of extensions of quantum
theory \cite{colbeck2011no}.

Consider a physical system prepared in a pure quantum state
$\ket{\psi}$, on which a measurement with setting $x$ is
performed, yielding an outcome $a$. Quantum theory provides
a probability distribution given by the Born rule: $p_{\mathrm{QM}}(a|x,\psi) =
\bra{\psi}\Pi_{a|x}\ket{\psi},$ where  $\Pi_{a|x}$ denotes the corresponding projector.

Suppose, however, that an extension of quantum theory
provides some supplementary information $\Xi$ that might
allow us to predict the outcome more accurately.
Such information need not take the form of classical
hidden variables, nor need it be described by quantum
theory: it could be encoded in an additional system of
any kind, whose measurement reveals information about
the outcome of the original experiment
\cite{colbeck2011no}.

To keep the description of $\Xi$ as general as possible,
Colbeck and Renner characterize it operationally through
the procedure by which it can be accessed. Let $z$ denote
the choice of measurement used to access $\Xi$, yielding
an outcome $c$. The supplementary information is assumed
to be accessible at arbitrary times and \emph{static},
meaning that the statistical behavior of its observation
does not depend on the spacetime coordinates at which
it is accessed.

\begin{figure}[h!]
    \centering
    \includegraphics[width=0.8\linewidth]{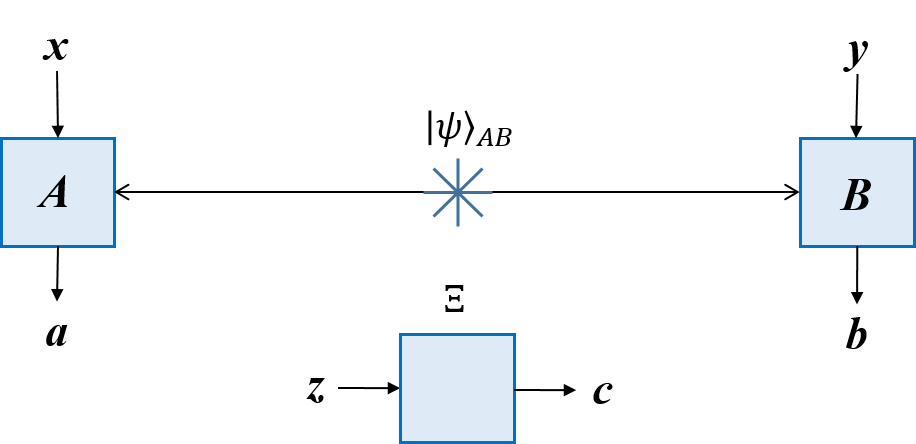}
    \caption{
Extended Bell-type scenario for the Colbeck--Renner
theorem. Alice and Bob share an entangled state
$\ket{\psi}_{AB}$ and perform local measurements
with settings $x$ and $y$, obtaining outcomes $a$
and $b$, respectively. Supplementary information
$\Xi$, not encapsulated by the quantum state $\ket{\psi}_{AB}$, can be accessed through an additional
measurement with setting $z$, yielding an outcome
$c$. Under the assumption that information is accessible and static, the three measurements can be considered in a configuration in which they are mutually spacelike separated.
}
    \label{cr}
\end{figure}

The question is whether knowing $c$, in addition to
the quantum state and the measurement setting, allows
a more accurate prediction of $a$ than quantum theory
alone, even without determining the outcome with certainty.

The first assumption is \emph{quantum-statistical agreement}:
the observed measurement statistics must agree with the
predictions of quantum theory
\begin{equation}
p(a|x,\psi)
=
p_{\mathrm{QM}}(a|x,\psi).
\label{crcompatibility}
\end{equation}
 The same requirement applies to
measurements on composite systems. In the following, the prepared quantum state is
understood to be fixed, and its dependence will
be omitted from the probability distributions.

Crucially, this assumption concerns the observed statistics
without conditioning on the supplementary outcome $c$.
If $c$ is recorded but not used, its possible values are
averaged over. Agreement with the Born rule therefore does
not, by itself, exclude the possibility that knowing $c$
allows more accurate predictions.

With reference to Fig.~\ref{cr}, consider now an extended
Bell-type scenario in which Alice and Bob share an
entangled state and perform local measurements with
settings $x$ and $y$, obtaining outcomes $a$ and $b$,
respectively. The supplementary information $\Xi$ can again
be accessed through an additional measurement with
setting $z$, yielding the outcome $c$.
The extended scenario is thus described by the joint
conditional probability distribution $p(a,b,c|x,y,z)$.

Colbeck and Renner consider a configuration in which
Alice's and Bob's measurements are spacelike separated.
Since $\Xi$ is assumed to be accessible at arbitrary
times and static, its observation can also be considered
spacelike separated from both measurements.
 In this configuration, the authors adopt a stronger
freedom-of-choice assumption than the one introduced
in Sect.~\ref{bellscenario}, which we shall call
\emph{spacetime freedom of choice} (called assumption $FR$ in Ref. \cite{colbeck2011no}).
This requires each freely chosen measurement setting
to be statistically independent of all other variables
whose associated spacetime events lie outside its
future light cone \cite{colbeck2011no}.

In the spacelike configuration considered here, this
requires
\begin{equation}
\begin{aligned}
p(x|y,z,b,c)&=p(x),\\
p(y|x,z,a,c)&=p(y),\\
p(z|x,y,a,b)&=p(z).
\end{aligned}
\label{crFR}
\end{equation}

Spacetime freedom of choice is stronger than ordinary
freedom of choice (Eq.~\eqref{freedom}), which requires
the measurement settings to be statistically independent
of the underlying hidden variables.\footnote{It is also
stronger than ordinary freedom of choice combined with
operational no-signaling (Eq.~\eqref{ns}). Indeed,
operational no-signaling does not exclude dependencies
on distant measurement settings that become apparent
when supplementary information is taken into account
\cite{leegwater2016impossibility}.}

The third assumption is the \emph{unitary description of quantum processes}: the quantum processes entering the argument, including measurements, admit a unitary description when, if necessary, the measuring apparatus and part of the environment are included. The statistics of subsequent measurements must agree with the quantum predictions for the resulting state \cite{colbeck2011no}. Importantly, Colbeck and Renner stress that this assumption ``need only hold for microscopic processes on short timescales and does not preclude subsequent wave function collapse" \cite{colbeck2011no}.

In Ref.~\cite{colbeck2011no}, the following result
was proven:

\begin{tcolorbox}[
    colback=paperblue!5,
    colframe=paperblue,
    sharp corners,
    boxrule=0.5pt
]

\textbf{Colbeck--Renner theorem}\\

Assuming (i) quantum-statistical agreement,
(ii) spacetime freedom of choice, and
(iii) the unitary description of quantum processes,
no extension of quantum theory can provide more
accurate predictions of measurement outcomes
on systems prepared in pure quantum states
by means of accessible and static supplementary
information.

\end{tcolorbox}
\begin{proof}

We sketch the argument of Ref.~\cite{colbeck2011no} for maximally entangled two-qubit states,
using the extended Bell-type scenario of Fig.~\ref{cr}.

Consider Alice and Bob sharing a Bell state, say
$\ket{\Phi^+}$. Quantum theory predicts that a
measurement of either qubit in any orthonormal
basis has two equally likely outcomes. Suppose,
however, that the supplementary information $\Xi$
allows Alice's outcome to be predicted more
accurately. Given some accessible supplementary
outcome $c$, the probabilities of Alice's two
outcomes would then differ from $1/2$. 

The spacetime freedom of choice assumption implies no-signaling constraints
on the extended distribution $p(a,b,c|x,y,z)$.
In particular, even when conditioned on $c$,
Alice's local outcome probabilities cannot
depend on Bob's measurement setting, and vice
versa. Ordinary no-signaling for the observed
outcomes $a$ and $b$ alone would not establish
this property.

Consider a sequence of local measurements on
the two qubits, with binary outcomes
$a,b\in\{-1,+1\}$ and settings
\[
x\in\{0,2,\ldots,2N-2\},
\qquad
y\in\{1,3,\ldots,2N-1\},
\]
where $N$ is a positive integer. Colbeck and Renner
consider the chained Bell expression\footnote{A closely related result, showing with the same chained Bell inequality that no-signaling eavesdroppers cannot predict the outcomes of measurements on maximally entangled states, had been obtained earlier in Ref.~\cite{barrett2006maximally}.}
\begin{equation}
I_N :=
\sum_{\substack{x,y\\|x-y|=1}}
p(a\neq b|x,y) + p(a=b|x=0,y=2N-1).
\label{crchain}
\end{equation}
Here the sum contains the probabilities that measurements with neighboring settings yield different outcomes,
while the last term is the probability that measurements at the two ends of the chain yield
equal outcomes. Bell-local correlations satisfy
the chained Bell inequality $I_N\geq1$. Indeed,
to minimize Eq. \eqref{crchain}, assume neighboring measurements always yielded equal
outcomes, the outcomes at the two ends would
necessarily be equal as well. Importantly, the following argument does not assume
Bell locality, but only the no-signaling constraints on the extended distribution established above.

For suitably chosen local measurements on the Bell
state, quantum theory instead predicts
\begin{equation}
I_N^{\mathrm{QM}}
=
2N\sin^2\left(\frac{\pi}{4N}\right)
\xrightarrow{N\to\infty}0.
\label{crquantumchain}
\end{equation}
For example, these correlations can be obtained
using measurement directions in the $xz$ plane,
with successive directions separated by an angle
$\pi/(2N)$. More generally, the measurement bases
can be chosen so that Alice's measurement at $x=0$
is any prescribed measurement and remains unchanged
as $N$ increases. The limit therefore constrains
the prediction of the outcome of that same measurement.

To see the role of these correlations, suppose
that, for a fixed supplementary outcome $c$,
Alice's measurement at $x=0$ is biased towards
$+1$. If Alice's and Bob's outcomes are strongly
correlated at neighboring settings, Bob's outcome
must have approximately the same bias. The
argument can be continued along the chain,
because no-signaling ensures that each party's
local outcome distribution, even when conditioned
on $c$, is independent of the distant setting.

The term closing the chain, however, requires
the outcomes at its two ends to be almost
perfectly anticorrelated. The successive
correlations therefore constrain Alice's
original outcome distribution to be close
to the distribution obtained by interchanging
$+1$ and $-1$. In particular, a nonzero bias
requires a nonzero value of the chained
expression conditioned on $c$.

The observed correlations are obtained by averaging
the correlations conditioned on $c$. For a fixed
choice $z$ of how to access the supplementary
information, the no-signaling constraints ensure
that the weights $p(c|z)$ do not depend on Alice's
and Bob's measurement settings. We may therefore
enlarge the measurement chain while keeping both
Alice's measurement at $x=0$ and the supplementary
observation fixed. Since every term in the chained
expression is nonnegative, the quantum prediction
$I_N\to0$ forces the conditioned chained expression
to approach zero for every supplementary outcome
$c$ with $p(c|z)>0$. The bound established in Appendix B
of Ref.~\cite{colbeck2011no} then implies
\begin{equation}
p(a=+1|x=0,z,c)=\frac12,
\label{crresult}
\end{equation}
and likewise for $a=-1$.

Thus, knowledge of $c$ leaves Alice's outcome
probabilities unchanged: access to $\Xi$ cannot
improve the quantum prediction. Since Alice's
measurement at $x=0$ was arbitrary, the conclusion
holds for any measurement. This completes the proof for maximally entangled two-qubit states.
 
Colbeck and Renner extend this result to arbitrary
measurements on pure quantum states using the
additional assumption concerning the unitary
description of quantum processes. They append
an auxiliary measurement that refines the original
outcomes into a larger set with probabilities
arbitrarily close to uniform. Using the unitary description of the enlarged measurement
process, the refined outcome can be recorded in an
apparatus that is arbitrarily close to maximally entangled with the remaining systems. The preceding argument can then be applied to the refined outcome.

Since the original measurement outcome can be
recovered from the refined outcome, supplementary
information cannot improve its prediction either.

\end{proof}

For maximally entangled states, this result generalizes earlier no-go results for specific nonlocal hidden-variable models in which the hidden variables bias the local outcome probabilities, most notably Leggett's model \cite{leggett2003nonlocal}, which had been tested and experimentally ruled out through so-called Leggett-type inequalities \cite{groblacher2007experimental,branciard2008testing} (see also the no-go theorem for covariant deterministic nonlocal hidden-variable extensions of quantum theory \cite{gisin2011impossibility}).

The Colbeck--Renner theorem has also prompted debate over the scope of its assumptions. The requirement that freely chosen settings satisfy the strong spacetime independence conditions used in the theorem has been questioned, since these conditions exclude hidden-variable models compatible with ordinary freedom of choice and operational
no-signaling \cite{ghirardi2013possible}. In response, it has been
argued that a freely chosen setting should be correlated
only with variables in its causal future
\cite{colbeck2013short}; moreover, their free-choice assumption matches Bell's own notion of ``free variables'', which he described as having ``implications only in their future light cones'' \cite{bell2001free}. The generalization to arbitrary 
measurements has also been questioned on the grounds
that it relies on an additional assumption about how
unitary evolution is represented in an extension of
quantum theory \cite{hermens2020completely}.

\subsection{Real versus complex quantum theory}

Quantum theory is conventionally formulated over complex Hilbert spaces,
but it is natural to ask whether complex numbers represent an essential
physical ingredient or merely a convenient mathematical tool. This question was recently addressed by Renou \emph{et al.}
\cite{renou2021quantum}, who showed that, in quantum networks
where independently prepared sources are represented by tensor-product
states, real and complex quantum theory can lead to experimentally
distinguishable predictions.\footnote{Within standard Bell scenarios, complex numbers can only be certified under an assumption on the dimension (see \cite{mosca}): for instance, correlations requiring complex (non-planar) qubit measurements have been observed experimentally \cite{christensen2015exploring}.}

By \emph{real quantum theory}, one means a theory that retains the Born
rule and the usual tensor-product composition of physical systems, but in
which states and measurement operators are defined on real Hilbert spaces of arbitrary dimensions.

\begin{figure}[h!]
    \centering
    \includegraphics[width=0.8\linewidth]{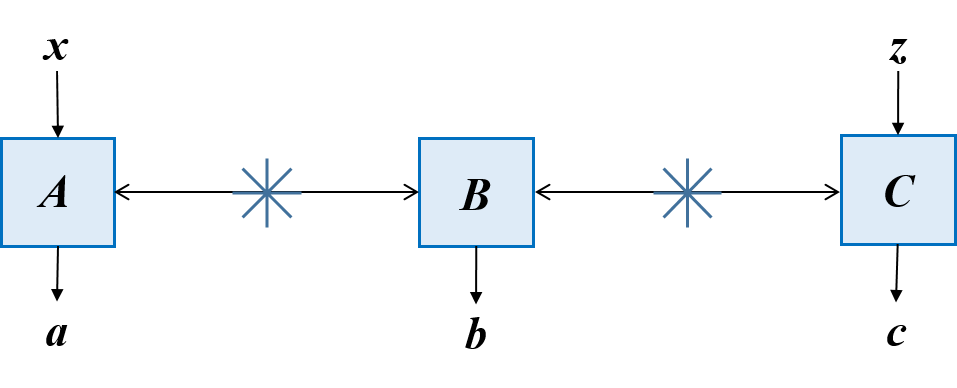}
    \caption{Bilocal quantum network consisting of two independent entangled-state sources and a central joint measurement. The resulting Alice–Charlie correlations, conditioned on Bob's outcome, provide an operational separation between real and complex quantum theory.}
    \label{imvsreal}
\end{figure}

Consider the entanglement-swapping network depicted in
Fig.~\ref{imvsreal}. Two independent sources each distribute an entangled
pair of qubits: the first pair is shared by Alice and Bob, while the second
is shared by Bob and Charlie. Alice and Charlie choose local measurements
labelled by $x$ and $z$, obtaining outcomes $a$ and $c$, respectively.
Bob performs a fixed joint measurement on the two qubits he receives and
obtains an outcome $b$.

Under the tensor-product representation of independent sources adopted in Ref.~\cite{renou2021quantum}, the global state factorizes as
\begin{equation}
\rho_{AB_1B_2C}
=
\rho_{AB_1}\otimes\rho_{B_2C}.
\end{equation}
The observed correlations are therefore described by
\begin{equation}
\begin{aligned}
p(a,b,c|x,z)
={}&
\operatorname{tr}\!\Big[
\big(
A_{a|x}\otimes B_b\otimes C_{c|z}
\big) \cdot
\\
&\qquad\quad
\big(
\rho_{AB_1}\otimes\rho_{B_2C}
\big)
\Big],
\end{aligned}
\end{equation}
where $A_{a|x}$ and $C_{c|z}$ are Alice's and Charlie's local measurement
operators, and $B_b$ is Bob's joint measurement operator on $B_1B_2$. In
a real quantum model, all states and measurement operators are represented
by real matrices, possibly acting on Hilbert spaces of arbitrary dimension.

Ref. \cite{renou2021quantum} proved the following result:

\begin{tcolorbox}[
    colback=paperblue!5,
    colframe=paperblue,
    sharp corners,
    boxrule=0.5pt
]
\textbf{Real-quantum-theory no-go theorem}\\

Any quantum model (i) based on real Hilbert spaces, (ii) in which probabilities are given by the Born rule, and (iii) in which independently prepared sources are represented by tensor-product states fails to reproduce certain correlations achievable in complex quantum theory.
\end{tcolorbox}

\begin{proof}
We provide a sketch of the proof based on the entanglement-swapping scenario of Fig.~\ref{imvsreal} \cite{renou2021quantum}. The argument exploits the quantum saturation of suitable Bell inequalities, not to rule out a class of classical models, as in Bell's theorem, but as a self-testing tool. Attaining their maximal quantum values constrains the underlying measurement operators.

We first describe the complex quantum realization. The two independent sources distribute two copies of the Bell state $\ket{\Phi^+}$, one between Alice and Bob and one between Bob and Charlie. Bob performs a Bell-state measurement on the two qubits he receives, yielding an outcome $b \in \{1, 2, 3, 4\}$. Conditioned on $b$, Alice and Charlie share one of the four maximally entangled Bell states.

Alice and Charlie choose inputs $x\in\{1,2,3\}$ and
$z\in\{1,\ldots,6\}$, respectively. In the complex quantum
realization, Alice measures
\begin{equation}
A_1=\sigma_z,\qquad
A_2=\sigma_x,\qquad
A_3=\sigma_y,
\end{equation}
while Charlie measures CHSH-like observables in each of the three
planes $x-z$, $y-z$, and $x-y$, i.e.,
\begin{equation}
\begin{aligned}
C_1&=\frac{\sigma_z+\sigma_x}{\sqrt{2}},
&
C_2&=\frac{\sigma_z-\sigma_x}{\sqrt{2}},
\\
C_3&=\frac{\sigma_z-\sigma_y}{\sqrt{2}},
&
C_4&=\frac{\sigma_z+\sigma_y}{\sqrt{2}},
\\
C_5&=\frac{\sigma_x-\sigma_y}{\sqrt{2}},
&
C_6&=\frac{\sigma_x+\sigma_y}{\sqrt{2}}.
\end{aligned}
\end{equation}
The corresponding outcomes are $a,c\in\{-1,+1\}$.

Bob's Bell-state measurement outcome $b$ determines which Bell
state is shared by Alice and Charlie. For each outcome $b$, define
the corresponding conditional correlators
\begin{equation}
\langle a_xc_z\rangle_b
:=
\sum_{a,c} ac\,p(a,c|x,z,b).
\end{equation}
Let $b_0$ denote the outcome for which Alice and Charlie share
$\ket{\Phi^+}$. For notational simplicity, in the following we write
$\langle a_xc_z\rangle:=\langle a_xc_z\rangle_{b_0}$.

For two inputs $x,x'$ of Alice and $z,z'$ of Charlie, define
\begin{equation}
\begin{aligned}
\mathrm{CHSH}(x,x';z,z')
:={}&
\langle a_xc_z\rangle
+\langle a_xc_{z'}\rangle
\\
&+\langle a_{x'}c_z\rangle
-\langle a_{x'}c_{z'}\rangle .
\end{aligned}
\end{equation}
Consider then the sum of three such CHSH expressions,
whose local bound is
\begin{equation}
\label{chsh3}
\begin{aligned}
\mathrm{CHSH3}:={}&
\mathrm{CHSH}(1,2;1,2)
+\mathrm{CHSH}(1,3;3,4)
\\
&+\mathrm{CHSH}(2,3;5,6)
\le 6.
\end{aligned}
\end{equation}
For the state $\ket{\Phi^+}$ and the measurements above, each
of the three CHSH expressions attains its maximal quantum value
$2\sqrt{2}$. Hence,
\begin{equation}
\label{maxval}
\mathrm{CHSH3}=6\sqrt{2}>6.
\end{equation}

The three CHSH tests share Alice's observables pairwise. Their simultaneous saturation therefore constrains the underlying quantum realization more strongly than any one of the three tests separately, a property that will be crucial in the argument below.

For each of the other outcomes of Bob's Bell-state measurement,
Alice and Charlie share a different Bell state. The corresponding
conditional correlations maximally violate a variant of
$\mathrm{CHSH3}$ obtained by simple relabelings of the measurement
outcomes.

To reach a contradiction, assume that the same correlations can be reproduced by a quantum model formulated over real Hilbert spaces of arbitrary dimension.

The maximal violation of each CHSH inequality constrains the corresponding observables to act as anticommuting Pauli operators on a maximally entangled qubit pair, up to local isometries and additional degrees of freedom. Since the three CHSH tests overlap, their simultaneous saturation imposes a common structure on Alice's three observables, corresponding to the three Pauli matrices $\sigma_x$, $\sigma_y$, and $\sigma_z$, up to complex conjugation. By imposing these constraints consistently for all four conditional states selected by Bob's measurement, Ref.~\cite{renou2021quantum} proves that any real-Hilbert-space realization of the full network correlations would require the Alice--Charlie marginal state before Bob's measurement not to admit a convex decomposition into products of real local states.

This is incompatible with source independence. Indeed, the
systems held by Alice and Charlie originate from two
independent sources, so their marginal state before Bob's
measurement is a product of real local states,
\begin{equation}
\rho_{AC}=\rho_A\otimes\rho_C,
\end{equation}
or, more generally, a convex combination of such products if shared randomness is allowed (as considered in Ref. \cite{renou2021quantum}).
Hence the correlations
predicted by the complex realization cannot be reproduced
by a real-Hilbert-space model satisfying the assumed
tensor-product representation of independent sources.\footnote{In a subsequent work it was shown that an analogous argument holds for a reduced number of settings, three for both Alice and Charlie \cite{bednorz2022optimal}.}
\end{proof}

The predictions of this no-go theorem have been experimentally confirmed in several independent experiments \cite{expreal1, expreal2, expreal3}, which observed violations of the bounds imposed by real-valued quantum theory.

The necessity of mathematically representing  source independence through tensor products, and more generally of the composition rule assumed for real quantum systems, has recently been questioned  \cite{hoffreumon2025quantum,hoffreumon2026falsified, barrios2026}, although it has also been shown that perfect independence is not required for the separation between real and complex quantum theory \cite{weilenmann2025partial}. The theorem remains valid for its explicitly stated composition rule, while the present debate concerns whether that rule is physically compelled by source independence \cite{moradikalarde2026comment}.

\subsection{Limits to agreement among quantum observers}

Over the last decade, Bell-like no-go theorems have also contributed to
advancing our understanding of the notorious quantum measurement problem. This becomes particularly striking when quantum
theory is applied not only to measured systems, but also to observers
performing measurements themselves, as illustrated by the \textit{Wigner's friend}
thought experiment \cite{wigner1961remarks}. An observer, usually called the
friend, measures a quantum system inside an isolated laboratory and obtains
a definite outcome. From the friend's perspective, the measurement has
therefore produced a single result. An external observer, Wigner, may instead
describe the entire laboratory --- including the friend and her memory ---
as evolving unitarily into an entangled superposition of different recorded
outcomes.

The ambiguity thus arises from two apparently legitimate quantum descriptions
of the same experiment: one in which the friend observes a definite outcome
and applies the ``collapse'' postulate, and another in which Wigner treats the
entire laboratory as an isolated quantum system evolving unitarily into an
entangled state between the friend's memory degree of freedom and the recorded
outcomes. Importantly, this tension cannot be dismissed as merely reflecting
Wigner's ignorance of what has occurred inside the laboratory. As pointed out
by Deutsch \cite{deutsch1985quantum}, the friend may communicate to Wigner that
a measurement has been performed and that a definite outcome has been
experienced, without revealing which outcome occurred. Since such a message
contains no information about the particular outcome, it can in principle be
encoded in a degree of freedom that remains uncorrelated with the outcome
record, allowing Wigner to receive this information while retaining a coherent
unitary description of the remaining degrees of freedom of the laboratory
\cite{brukner2018no, baumann2020wigner, delsanto2025conditioning}.

In recent years, extended Wigner's-friend scenarios combining multiple distant
observers with Bell-type arguments have turned this tension into precise no-go
theorems. Rather than merely contrasting collapse with unitary evolution, these
results ask whether the quantum predictions associated with Wigner's description
can be consistently combined with assumptions concerning the definiteness,
observer-independence, and correlations of measurement outcomes.

One way to evade such theorems is to deny the universal applicability of
unitary quantum theory, as in objective-collapse models. In that case, once a
physical collapse occurs, the coherent description of the entire laboratory
required by Wigner is no longer valid. In the following, we instead consider
the consequences of retaining the quantum description of the laboratory while
imposing additional assumptions on measurement outcomes.

Brukner considered a Bell-type scenario involving two distant Wigner's-friend laboratories and showed that locality, freedom of choice,
the universal validity of quantum theory,\footnote{Here universal validity means that ``Quantum predictions hold at any scale, even if the measured system contains objects as large as an `observer' (including her laboratory, memory etc.)"
\cite{brukner2018no}} and the existence of
observer-independent facts cannot all be maintained simultaneously
\cite{brukner2016quantum}. Unlike in standard Bell tests, where incompatible (noncommuting) local observables cannot be measured jointly, Brukner's setup allows for a joint measurement of all relevant observables for a particular choice of the Wigners' measurement settings. The resulting joint probability distribution is then assumed to underlie the statistics of other measurement settings as well, even though not all relevant observables are actually measured in those settings; this assumption led to a critical debate \cite{brukner2018no, healey2018quantum, baumann2019comment, healey2019reply}.

A different extension was introduced by Frauchiger and Renner \cite{frauchiger2018quantum}. Their argument considers multiple agents who use quantum theory to reason about measurements performed by other agents. They showed that the following three assumptions are mutually incompatible: (i) the universal applicability of the quantum-mechanical Born rule, according to which an agent may regard a proposition as certain whenever quantum theory assigns it probability one, also when the system under consideration includes other agents and their laboratories; (ii) consistency among the conclusions drawn by different agents; and (iii) the occurrence of a single outcome for every measurement. Taken together, these assumptions lead to a contradiction, and hence their conjunction cannot be consistently upheld.

\begin{figure}[h!]
    \centering
    \includegraphics[width=0.9\linewidth]{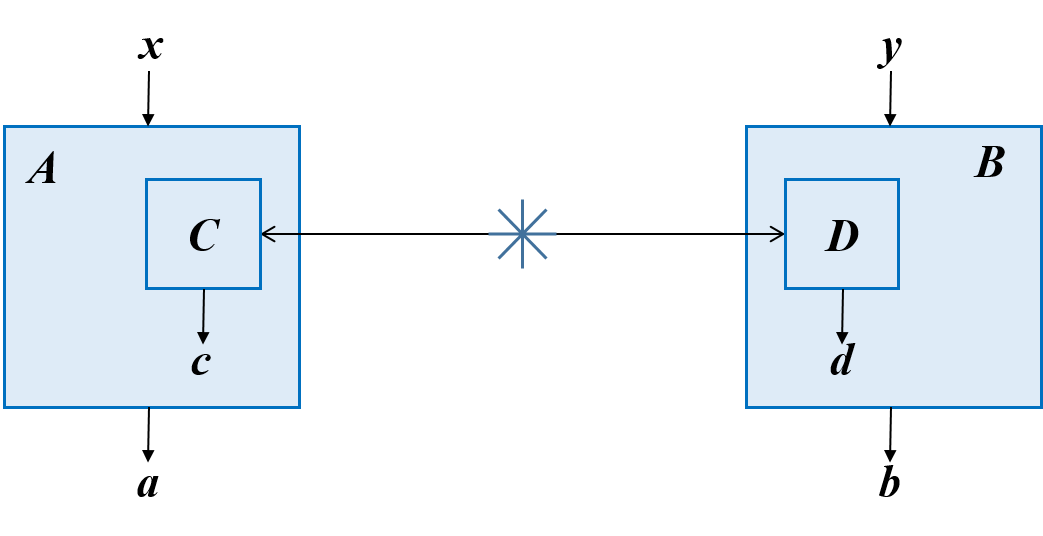}
    \caption{
   Extended Wigner's-friend local-friendliness scenario.
Charlie and Debbie each receive one half of a partially entangled state
and perform a fixed measurement inside their respective laboratories,
recording outcomes $c$ and $d$. The superobservers Alice and Bob
subsequently choose measurements on the corresponding laboratories,
obtaining outcomes $a$ and $b$. One setting corresponds to reading the
friends' records, while the remaining settings involve coherently
reversing the friends' measurements and performing alternative
measurements on the recovered qubits.
    }
    \label{friendliness}
\end{figure}

\begin{figure}[h!]
    
    \includegraphics[width=0.9\linewidth]{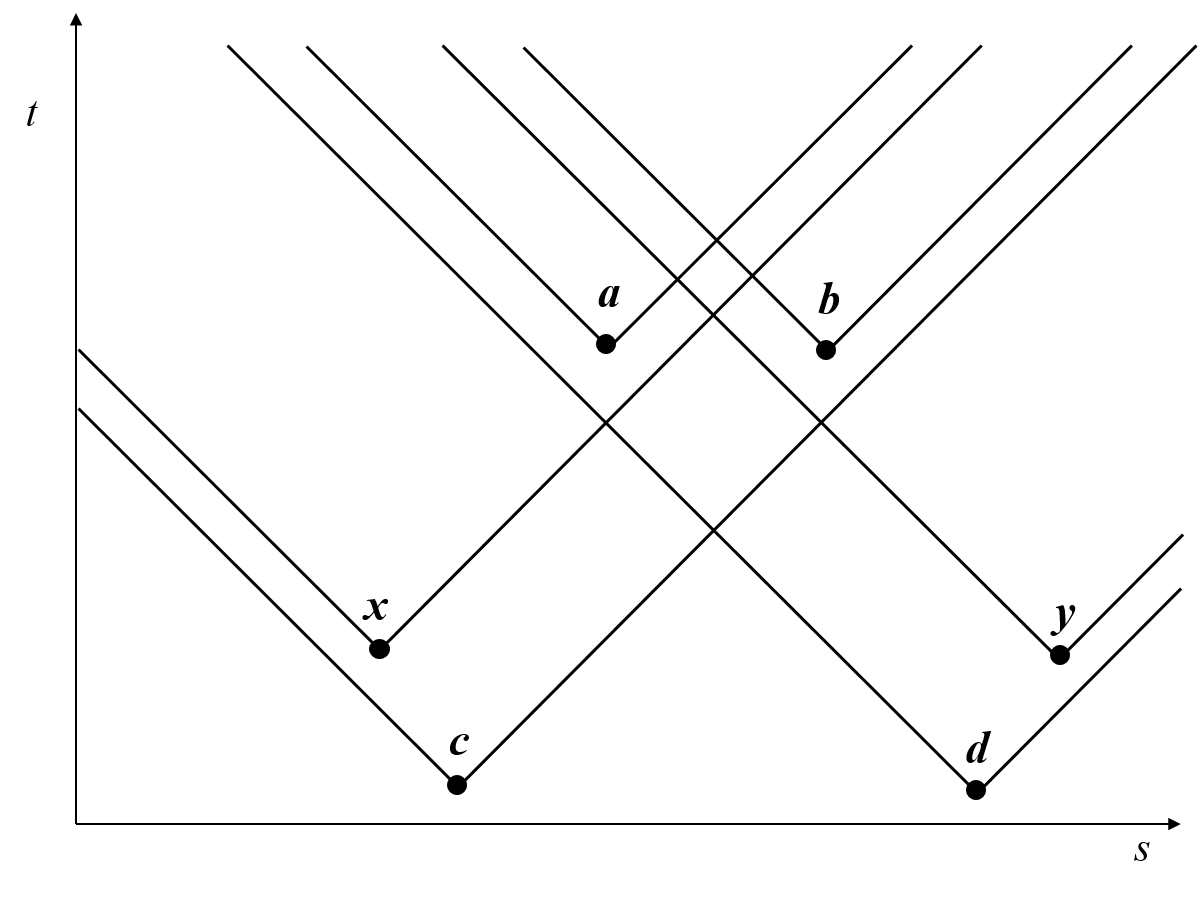}
    \caption{Spacetime diagram of the inputs and outputs of the local friendliness scenario (see main text). To avoid confusion with Alice's setting $x$, the spatial coordinate is here denoted by $s$.}
    \label{lfcausalgraph}
\end{figure}

Building on Brukner's scenario, Bong \emph{et al.} derived the
local-friendliness no-go theorem \cite{bong2020strong}. In the extended Wigner's-friend experiment
depicted in Fig.~\ref{friendliness}, Charlie and
Debbie each measure one half of an entangled state inside isolated
laboratories and record outcomes $c$ and $d$. Alice and Bob subsequently
choose measurements $x$ and $y$ on the corresponding laboratories,
obtaining outcomes $a$ and $b$. For $x=1$, Alice reads Charlie's record,
and  for $y=1$, Bob reads Debbie's record. For the remaining settings,
Alice and Bob instead coherently reverse the friends' measurement
interactions and perform alternative measurements on the recovered
quantum systems. The spacetime diagram of this scenario is depicted in Fig. \ref{lfcausalgraph}.

Bong \textit{et al.} derive a theorem based on the following assumptions \cite{bong2020strong}: (i) absoluteness of observed events, (ii) no-superdeterminism, and (iii) locality of  observed events.
 The conjunction of these assumptions, which we now define more formally, is called \emph{local friendliness}.

First, \emph{absoluteness of observed events} states that every observed
event is a single event that exists absolutely, rather than relative to a
particular observer.\footnote{A closely related notion, \emph{observer-independent facts}, was considered in Refs.~\cite{brukner2016quantum,brukner2018no}.
We retain the original terminology to avoid inconsistency with the existing literature.} 

In the present scenario, the absoluteness of observed events requires the existence of a joint probability distribution $p(a,b,c,d|x,y)$ such that the empirical correlations observed by Alice and Bob satisfy
\begin{equation}
p(a,b|x,y)
=
\sum_{c,d} p(a,b,c,d|x,y),
\end{equation}
and such that Alice and Bob agree with the corresponding friends whenever
they read their records,
\begin{equation}
\begin{aligned}
p(a|c,d,x=1,y)&=\delta_{a,c},\\
p(b|c,d,x,y=1)&=\delta_{b,d}.
\end{aligned}
\end{equation}
Importantly, this assumption does not assign outcomes to measurements that
were not performed. It requires only that all actually observed outcomes
$a,b,c,d$ possess observer-independent values.

Second, \emph{no-superdeterminism} is the condition
\begin{equation}
\label{nosupdet}
p(x,y|c,d)=p(x,y),
\end{equation}
which captures the assumption that the measurement settings chosen by
Alice and Bob are statistically independent of the earlier outcomes
recorded by Charlie and Debbie.\footnote{Note that this condition
is mathematically equivalent to $p(c,d|x,y)=p(c,d)$, which is the form
used in the original paper \cite{bong2020strong}. Given the temporal
ordering of the operations, we find Eq.~\eqref{nosupdet} more intuitive.}

Third, \emph{locality of observed events} reads, in the present scenario as
\begin{equation}
\begin{aligned}
p(a|c,d,x,y)&=p(a|c,d,x),\\
p(b|c,d,x,y)&=p(b|c,d,y).
\end{aligned}
\end{equation}
This requires that the probability of
an observed event be unaffected by a freely chosen setting at spacelike separation, even when conditioned on other events lying outside the future light cone of that choice.\footnote{Note that in Ref.~\cite{bong2020strong} this assumption is simply called ``locality.'' This terminology is potentially misleading in the present context, since the condition is not Bell locality as defined in Eq.~\eqref{belllocal}, but requires only that the probability of an observed event be unaffected by a freely chosen setting at spacelike separation. We shall therefore refer to it more explicitly as \emph{locality of observed events}. Moreover, although these conditions resemble no-signaling, they concern probabilities conditioned on the friends' outcomes $c,d$. Together with no-superdeterminism, Eq.~\eqref{nosupdet}, they imply the usual operational no-signaling conditions for Alice's and Bob's observed marginals.}

Based on the previous assumptions, it is possible to prove the following \cite{bong2020strong}:

\begin{tcolorbox}[
    colback=paperblue!5,
    colframe=paperblue,
    sharp corners,
    boxrule=0.5pt
]
\textbf{Local-friendliness no-go theorem}\\

Assuming the universal applicability of unitary quantum theory, certain quantum correlations predicted in extended Wigner’s-friend scenarios cannot be reproduced by any theory satisfying local friendliness (i.e., the conjunction of absoluteness of observed events, no-superdeterminism, and locality of observed events).

\end{tcolorbox}
\begin{proof}
    We provide a sketch of the argument for the simplest case in which Alice and Bob each have three measurement settings, $x,y\in\{1,2,3\}$, with binary outcomes $a,b\in\{-1,+1\}$. For $x=1$, Alice reads Charlie's measurement record, so that $a=c$, while for $y=1$, Bob reads Debbie's record, so that $b=d$. For $x\in\{2,3\}$ and $y\in\{2,3\}$, Alice and Bob instead coherently reverse the corresponding friend's measurement interaction and subsequently perform an alternative projective measurement on the recovered qubit.

The assumptions of local friendliness constrain the observable correlations $p(a,b|x,y)$ to belong to the local-friendliness polytope.\footnote{Local-friendliness polytopes were previously derived in a different context and known as \textit{partially deterministic polytopes} \cite{woodhead2014imperfections}.} For two measurement settings per superobserver, the local-friendliness set and the Bell-local set coincide, but already for three binary settings the Bell-local polytope is strictly contained in the local-friendliness polytope \cite{bong2020strong},
\begin{equation}
\mathcal{L}_{\mathrm{Bell}}
\subsetneq
\mathcal{L}_{\mathrm{LF}} .
\end{equation}
Thus, local friendliness is a weaker requirement than Bell locality: some correlations that are Bell nonlocal are nevertheless compatible with all the local-friendliness assumptions.

One can then derive genuine local-friendliness inequalities for this scenario \cite{bong2020strong}, e.g.,
\begin{equation}
\begin{aligned}
I_{\mathrm{LF}}
:={}&
-\langle a_1\rangle
-\langle a_2\rangle
-\langle b_1\rangle
-\langle b_2\rangle
\\
&-\langle a_1 b_1\rangle
-2\langle a_1 b_2\rangle
-2\langle a_2 b_1\rangle
+2\langle a_2 b_2\rangle
\\
&-\langle a_2 b_3\rangle
-\langle a_3 b_2\rangle
-\langle a_3 b_3\rangle
\leq 6 .
\end{aligned}
\end{equation}
Since the Bell-local polytope is strictly contained in the local-friendliness polytope, violating this inequality rules out models based on the assumptions of local friendliness, rather than merely Bell locality.

Quantum mechanics predicts correlations that violate this bound. For a suitable partially entangled two-qubit state and projective measurements, found by numerical optimization, one obtains \cite{bong2020strong}
\begin{equation}
I_{\mathrm{LF}}\simeq 7.345>6,
\end{equation}
which semidefinite programming shows to be the maximal quantum value.
The corresponding quantum correlations therefore lie outside the local-friendliness polytope. Hence, absoluteness of observed events, no-superdeterminism, and locality of observed events cannot all be simultaneously maintained if the corresponding predictions of quantum theory are to be reproduced,  even allowing for Bell nonlocality.
\end{proof}

Conclusive experimental tests of extended Wigner's friend scenarios involving genuine observers are, however, still far beyond experimental reach (if they would ever become feasible). The main, highly nontrivial issue is what physical system should count as an observer in such experiments. Existing proof-of-principle experiments in extended Wigner's-friend scenarios
\cite{proietti2019experimental,bong2020strong} represent the friends by microscopic
quantum systems, whose status as genuine observers can naturally be questioned.
To address this issue, Wiseman \textit{et al.} proposed a ``thoughtful'' version of
the local-friendliness theorem in which the friend is a ``human-level artificial
intelligence'' implemented reversibly on a sufficiently powerful quantum computer
\cite{wiseman2023thoughtful}. This proposal introduces the further debatable
assumption that such an artificial intelligence should in principle qualify as a genuine
\textit{observer}. Moreover, such an experiment remains, for the foreseeable future, in
the realm of science fiction. Nevertheless, these proposals make explicit the
ambiguity surrounding what should count as an observer in Wigner's-friend tests. 

Moreover, extended Wigner's-friend scenarios have led to additional no-go theorems constraining possible descriptions of quantum observers and their measurement outcomes \cite{healey2018quantum,zukowski2021physics,guerin2021nogo, baumann2024memory,haddara2023possibilistic}. Other works have clarified how different observers should (dis)agree in assigning quantum states  \cite{baumann2020wigner,delsanto2025conditioning, diaz2025quantum}, as well as the logical structure of multi-agent paradoxes and their extension beyond quantum theory \cite{nurgalieva2019inadequacy,vilasini2019multiagent}.


\begin{table*}[t]

\centering
\renewcommand{\arraystretch}{1}
\setlength{\tabcolsep}{4pt}

\resizebox{\textwidth}{!}{%
\begin{tabular}{|c|c|c|c|}
\hline

\textbf{No-go theorem}
&
\textbf{Minimal scenario}
&
\textbf{Main assumptions}
&
\textbf{Excluded models}
\\
\hline

\tcell{3.2cm}{0.9cm}{
Bell's theorem
}
&
\tcell{4.5cm}{0.9cm}{
Bipartite Bell scenario
}
&
\tcell{4.5cm}{0.9cm}{
Locality and freedom of choice
}
&
\tcell{4.5cm}{0.9cm}{
Local hidden-variable models
}
\\
\hline
\hline

\tcell{3.2cm}{1.35cm}{
Nonlocality without inputs theorem 
}
&
\tcell{4.5cm}{1.35cm}{
Triangle network with three independent sources and three fixed joint measurements
}
&
\tcell{4.5cm}{1.35cm}{
Locality and source independence
}
&
\tcell{4.5cm}{1.35cm}{
Triangle-local classical models
}
\\
\hline

\tcell{3.2cm}{1.35cm}{
Finite-speed superluminal-influence no-go theorem
}
&
\tcell{4.5cm}{1.35cm}{
Three-partite spacetime scenario
}
&
\tcell{4.5cm}{1.35cm}{
Finite-speed superluminal influences in a privileged reference frame
and operational no-signaling
}
&
\tcell{4.5cm}{1.35cm}{
Operationally no-signaling \(v\)-causal models
}
\\
\hline

\tcell{3.2cm}{1.35cm}{
PBR theorem
}
&
\tcell{4.5cm}{1.35cm}{
Two independent preparations and one joint measurement
}
&
\tcell{4.5cm}{1.35cm}{
Ontological-model framework and preparation independence
}
&
\tcell{4.5cm}{1.35cm}{
Preparation-independent \(\psi\)-epistemic models
}
\\
\hline

\tcell{3.2cm}{1.35cm}{
Colbeck--Renner theorem
}
&
\tcell{4.5cm}{1.35cm}{
Bipartite Bell scenario with accessible supplementary information
}
&
\tcell{4.5cm}{1.35cm}{
Quantum-statistical agreement,
spacetime freedom of choice,
and unitary quantum dynamics
}
&
\tcell{4.5cm}{1.35cm}{
Extensions of quantum theory with improved predictive power
}
\\
\hline

\tcell{3.2cm}{1.35cm}{
Real-quantum-theory no-go theorem
}
&
\tcell{4.5cm}{1.35cm}{
Bilocal network with two independent sources and one joint measurement
}
&
\tcell{4.5cm}{1.35cm}{
Real Hilbert spaces, the Born rule, and tensor-product representation
of independent sources
}
&
\tcell{4.5cm}{1.35cm}{
Real-Hilbert-space quantum models satisfying the tensor product composition rule
}
\\
\hline

\tcell{3.2cm}{1.35cm}{
Local-friendliness no-go theorem
}
&
\tcell{4.5cm}{1.35cm}{
Extended Wigner's-friend-Bell scenario with two pairs of nested observers
}
&
\tcell{4.5cm}{1.35cm}{
Absoluteness of observed events, no-superdeterminism,
and locality of observed events
}
&
\tcell{4.5cm}{1.35cm}{
Local observer-independent models
}
\\
\hline

\end{tabular}%
}

\caption{Comparison between Bell's theorem and the no-go theorems in
extended  scenarios discussed in this review.}
\label{tab:no-go-theorems}

\end{table*}


\section{Conclusions and outlook}

We have reviewed a selection of no-go theorems that, in our view,
significantly extend the conceptual legacy of Bell's theorem. As summarized
in Table~\ref{tab:no-go-theorems}, these results use increasingly
sophisticated operational scenarios to constrain possible explanations of
quantum phenomena, including local models in networks
without measurement inputs, finite-speed superluminal influences,
epistemic interpretations of the quantum state, the predictive power of supplementary information,  formulations of quantum theory over real Hilbert spaces, and observer-independent accounts of measurement outcomes.

The results discussed here are by no means intended to provide an exhaustive
account of the many developments in quantum foundations since the time of  Bell's theorem. Several other
operational arguments have established important constraints on the
structure and interpretation of quantum theory.

For instance, Wood and Spekkens established a further general limitation on classical causal explanations of Bell-violating correlations: They showed that any classical causal model reproducing both a Bell violation and the observed no-signaling independences must rely on fine-tuning of its causal parameters  \cite{wood2015lesson}. More recently,
a multipartite extension of Reichenbach's common-cause principle has shown
that, within quantum theory, perfect randomized coordination among several
parties requires a common cause shared by all of them
\cite{centeno2026coordination}. 

Moreover, Bell's theorem shows that quantum correlations cannot be reproduced without communication, but one may still ask how much communication would suffice \cite{toner2003communication}. For two entangled qubits, two classical bits are enough \cite{renner2023classical}, yet for two entangled four-dimensional systems no bounded amount of classical communication can exactly reproduce all quantum correlations, even with unlimited shared randomness \cite{degois2026quantum}. Nevertheless, exact simulation is possible if one allows for a finite average number of bits of communication, even though, with nonzero probability, an arbitrarily large number of bits is communicated \cite{castellano2026cost}. In more general scenarios, however, where the transmitted system can be jointly measured with other quantum systems held by the receiver, even a single qubit cannot be simulated with finite classical communication \cite{gopalkrishna2026no}.

Further developments in quantum networks have shown that states that are Bell-local at the single-copy level may become nonlocal when several copies are considered jointly, even without assuming state independence
\cite{cavalcanti2011quantum,palazuelos2012superactivation}.
Moreover, any connected network in which the parties are linked by bipartite pure entangled states can generate genuinely multipartite nonlocal correlations
\cite{contrerastejada2021genuine}.

A further development in the foundations of measurement theory concerns the question of which joint measurements on spatially separated quantum systems are physically realizable without violating no-signaling. This problem was first highlighted in quantum field theory by Sorkin, who showed that the usual ideal-measurement state-update rule can enable superluminal signaling despite microcausality, leading to what he called “impossible measurements” \cite{sorkin1993impossible}. Independently, Popescu and Vaidman derived strong causality constraints on nonlocal measurements, showing that an ideal, i.e. immediately repeatable, verification measurement of an entangled state is possible only if all locally accessible information about the initial state is erased \cite{popescu1994causality}. Nevertheless, once ideality of measurements is relaxed, arbitrary nonlocal measurements can in principle be localized using shared entanglement and joint measurements in a network scenario, as shown by Vaidman \cite{vaidman2003instantaneous}, thereby suggesting a possible resolution of Sorkin’s impossible-measurement problem \cite{gisin2024towards}. This has recently motivated a classification of joint measurements according to their entanglement cost of localization, directly connecting this foundational question with quantum networks \cite{pauwels2025classification}. Several approaches have been proposed to provide quantum field theory with a fully consistent measurement framework that overcomes the problem of impossible measurements \cite{fewster2020quantum,borsten2021impossible}, but the status and interpretation of such constructions remain the subject of active foundational debate \cite{mandrysch2026causality}.

Another extension of Bell-like reasoning concerns causal order. Several operational frameworks have been developed to describe quantum processes with nontrivial temporal and causal structure \cite{taranto2025higher}, including quantum combs \cite{chiribella2009theoretical} and the process-matrix framework \cite{oreshkov2012quantum}. In particular, for fixed finite numbers of parties, inputs, and outputs, correlations compatible with a definite, possibly probabilistic, causal order form a convex polytope whose facets define causal inequalities, in close analogy with Bell inequalities \cite{branciard2016simplest}. A Bell theorem has also been formulated for temporal order itself, which can become entangled when it depends on a massive body in spatial superposition \cite{zych2019bell}. The physical realizability of processes violating causal inequalities and, more generally, the interpretation, dynamics, and device-independent certification of indefinite causal structures remain subjects of ongoing debate  \cite{castro2018dynamics, oreshkov2019time, purves2021quantum, camati2023comparing, van2023device, vilasiniPRA, vilasiniPRL,  taranto2025higher}. 

Finally, recent no-go results in the spirit of Bell's theorem, which are attracting increasing attention, concern the quantum nature of gravity. Proposals to witness the nonclassicality of gravity through gravitationally induced entanglement were put forward independently in Refs.  \cite{bose2017spin,marletto2017gravitationally}. More general results subsequently showed that, under suitable
assumptions about mediation and locality, the generation of entanglement through gravity is incompatible with a classical description of the gravitational mediator (see, e.g., \cite{marletto2020witnessing, galley2022nogo}) and that any consistent coupling between classical gravity and quantum matter must be irreversible \cite{galley2023any}.

Taken together, these developments suggest that Bell's theorem initiated not merely the study of quantum nonlocality, but a broader methodology for identifying the limits of physical descriptions, and even metaphysical assumptions, through operationally testable impossibility results. As increasingly complex quantum systems, networks, observers, causal structures, and gravitational degrees of freedom become subject to coherent experimental control, this methodology is likely to play an increasingly central role in quantum foundations.

We conclude with the hope --- and in fact the admonition --- that foundational research of the kind reviewed here, which only emerged from the margins of physics after considerable struggle \cite{junior2014quantum}, will not once again be shortsightedly left behind in the growing emphasis on applications and technological development.

\begin{acknowledgments}
We thank Antonio Acín, Jean-Daniel Bancal,  \v Caslav Brukner,  Riccardo Castellano, Flaminia Giacomini, Yeong-Cherng Liang, Alejandro Pozas-Kerstjens, Renato Renner, Valerio Scarani, and Mirjam Weilenmann for useful comments and for pointing out relevant literature.
\end{acknowledgments}



\bibliography{bib}

\end{document}